\documentclass[11pt]{article}
\pdfoutput=1
\usepackage{jheppub}
\usepackage[T1]{fontenc}
\usepackage{color}
\usepackage{float}
\usepackage{url}
\usepackage{braket}
\usepackage{amsmath}
\usepackage{amsthm}
\usepackage{amssymb}
\usepackage{amsfonts}
\usepackage{graphicx}
\usepackage{caption}
\usepackage{subcaption}
\usepackage{multirow,array}
\usepackage{qcircuit}
\usepackage{hyperref}
\usepackage{comment}
\definecolor{red}{rgb}{1,0.,0}

\usepackage[mathscr]{eucal}
\usepackage[bbgreekl]{mathbbol}
\usepackage{mathtools}
\usepackage{tikz}
\usepackage[compat=1.1.0]{tikz-feynman}
\usepackage{feynmf}

\newcommand{\be}{\begin{equation}}
\newcommand{\ee}{\end{equation}}
\newcommand{\bea}{\begin{eqnarray}}
\newcommand{\eea}{\end{eqnarray}}

\def\({\left(}
\def\){\right)}
\def\[{\left[}
\def\]{\right]}

\let\ev=\bracket

\def\ap{{\alpha'}}

\usepackage{xcolor}
\colorlet{myPurple}{blue!40!red}

\usepackage{color}

\subheader{\begin{flushright}
\texttt{IFT-UAM/CSIC-22-138}
\end{flushright}}

\title{Worldsheet traversable wormholes}

\author[a]{Jan de Boer,}
\author[b]{Viktor Jahnke,}
\author[b]{Keun-Young Kim}
\author[c]{and Juan F. Pedraza}

\affiliation[a]{Institute for Theoretical Physics, University of Amsterdam, Amsterdam, The Netherlands}
\affiliation[b]{School of Physics and Chemistry, Gwangju Institute of Science and Technology, Gwangju, Korea}
\affiliation[c]{Instituto de F\'isica Te\'orica UAM/CSIC, Calle Nicol\'as Cabrera 13-15, Madrid 28049, Spain}

\emailAdd{j.deboer@uva.nl}
\emailAdd{viktorjahnke@gist.ac.kr}
\emailAdd{fortoe@gist.ac.kr}
\emailAdd{j.pedraza@csic.es}

\abstract{
We construct \emph{worldsheet traversable wormholes} by considering the effects of a double-trace deformation, $\delta\mathcal{L}\sim h\partial\phi_L\partial\phi_R$, coupling the endpoints of an open string in AdS space. The operator deforming the theory is irrelevant and makes the boundaries bend inward toward the IR. This effect, reminiscent of two-dimensional dilaton gravities, renders the teleportation protocol more efficient and facilitates the transfer of information between the members of the dual Bell pair. We compare our results with those obtained with the standard double-trace deformation, $\delta\mathcal{L}\sim h\phi_L\phi_R$, introduced by Gao, Jafferis and Wall.}

\begin{document}
\maketitle

\section{Introduction}

The landscape of spacetimes that could emerge from Einstein's field equations began to take on a variety of novel configurations not long after general relativity was discovered.
A noteworthy example is a solution discovered in the mid '30s by Einstein himself and Rosen, describing a \emph{bridge between worlds}, and
known thereafter under the name of Einstein-Rosen bridge \cite{Einstein:1935tc}. An Einstein-Rosen bridge is a vacuum solution to Einstein's equations that link two otherwise spatially disconnected spacetimes, what in modern terminology is referred to as a maximally extended Schwarzschild solution. Many generalizations have appeared since, including bridges connecting distant but causally connected regions of the same geometry and, thus, representing actual \emph{shortcuts} in spacetime. These solutions were popularized after a seminal paper by Misner and Wheeler in the late '50s \cite{Misner:1957mt}, which coined the term \emph{wormholes}.

A crucial property of these wormholes is that energy-violating matter is typically needed to support their throats, which makes them seem to belong only in science fiction. However, recent progress has shown that these violations may naturally occur in quantum mechanics, suggesting that real-world traversable wormholes might exist. This raises fascinating questions about their nature and phenomenology, with some even proposing that they may be created in a laboratory, and use them in experiments of quantum teleportation \cite{Brown:2019hmk, Nezami:2021yaq}.

The interest in wormhole solutions was reignited with the discovery of the AdS/CFT correspondence \cite{Maldacena:1997re}, a robust framework that allows us to pose questions about quantum gravity and reformulate them in the language of field theory. In this context, a maximally extended or two-sided AdS black hole is known to map to the canonical purification of the thermal density matrix, known as the thermofield double state \cite{Maldacena:2001kr}. This unique, maximally entangled state is thus represented by an Einstein-Rosen bridge embedded in AdS space, leading to one of the central theses of the AdS/CFT correspondence, namely, the fact that entanglement generates connectivity in the bulk spacetime \cite{VanRaamsdonk:2010pw}. The ER=EPR correspondence, which connects EPR-type entanglement in boundary theory with the presence of an Einstein-Rosen bridge in dual bulk spacetime, provides a qualitative yet catchy formulation of this claim.  Evidence for this correspondence includes \cite{Jensen:2013ora,Sonner:2013mba,Chernicoff:2013iga,Jensen:2014bpa,Fischler:2014ama}, showing that single Bell pairs in the field theory (EPR or Hawking) can be thought of as generating Planckian-size wormholes, holographically represented by strings in AdS with an Einstein-Rosen bridge in the induced geometry.

Einstein-Rosen bridges in AdS are, however, not traversable. This is consistent with the fact that while entangled, the theories in the thermofield double state do not interact. Traversable wormholes, on the other hand, require a violation of the so-called Average Null Energy Condition (ANEC). For a local quantum field theory, the ANEC states that the integral of the stress-energy tensor along a complete achronal null geodesic is non-negative
\begin{equation} \label{eq-ANEC}
\int T_{\mu \nu} k^{\mu} k^{\nu} d \lambda \geq 0\,,
\end{equation}
where $k^{\mu}$ is the tangent vector and $\lambda$ is an affine parameter. Classically, a violation of ANEC is prevented by the Null energy Condition (NEC), $T_{\mu \nu} k^{\mu} k^{\nu} \geq 0$, which must be valid for any physically reasonable theory, and implies (\ref{eq-ANEC}). However, quantum mechanical effects may induce negative null energy, leading to violations of the NEC and the ANEC. A prototypical example where such violations may be triggered was studied by Gao, Jafferis and Wall (GJW) \cite{Gao:2016bin}, which showed that a double trace deformation of the form
\begin{equation} \label{eq-deformationGJW}
    \delta H = \int dx\, h(t,x)\,\mathcal{O}_L(-t,x)\, \mathcal{O}_R(t,x)\,,
\end{equation}
introduces the negative null energy in the bulk necessary to sustain a traversable wormhole.\footnote{Technically, however, the coupling (\ref{eq-deformationGJW}) breaks one of the assumptions of the proof of ANEC, since it is non-local. This deformation renders all null geodesics connecting the two sides of an eternal AdS black hole \emph{non-achronal}.} They worked within the semi-classical approximation, where matter fields are treated quantum mechanically, but the gravitational field, or the metric, is treated classically. In this context, one is to solve the semi-classical Einstein's equations
\begin{equation}
    G_{\mu \nu} = 8 \pi G_N \langle T_{\mu \nu} \rangle\,,
\end{equation}
where $G_{\mu \nu}$ is Einstein's tensor, and $\langle T_{\mu \nu} \rangle\,$ is the expectation value of the stress-energy tensor in a given quantum state. GJW showed that for certain choices of the coupling $h(t,x)$, the 1-loop expectation value of the stress-energy tensor does indeed lead to
\begin{equation}
    \int \langle T_{\mu \nu}\rangle k^{\mu} k^{\nu} d\lambda <0\,,
\end{equation}
rendering an Einstein-Rosen bridge traversable. This allows the transfer of information between the two asymptotic boundaries, a process that can be viewed as a teleportation protocol~\cite{Gao:2016bin}. GJW considered the specific case of a two-sided BTZ black hole, but their construction was subsequently extended in several directions, for example, to black holes solutions in JT gravity~\cite{Maldacena:2017axo,Bak:2018txn,Bak:2019mjd}, higher-dimensional black holes~\cite{Maldacena:2018gjk, Ahn:2020csv,Ahn:2022rle}, asymptotically flat black holes~\cite{Fu:2019vco}, multi-boundary traversable wormholes~\cite{AlBalushi:2020kso,Emparan:2020ldj}, eternal traversable wormholes~\cite{{Maldacena:2018lmt,Freivogel:2019lej,Bintanja:2021xfs}}, near-extremal black holes~\cite{Fallows:2020ugr}, and rotating black holes~\cite{Caceres:2018ehr}. Other interesting developments include for instance~\cite{Maldacena:2020sxe,Almheiri:2018ijj, Bao:2018msr, Freivogel:2019whb, Couch:2019zni, Marolf:2019ojx,Fu:2018oaq, Hirano:2019ugo, Freivogel:2021ivu,Geng:2020kxh, Nosaka:2020nuk,Levine:2020upy,Garcia-Garcia:2019poj,Numasawa:2020sty, Anand:2020wlk,Anand:2022sbd}. For a recent review of wormholes in holography, see \cite{Kundu:2021nwp}.

The stringy setups considered in \cite{Jensen:2013ora,Sonner:2013mba,Chernicoff:2013iga,Jensen:2014bpa,Fischler:2014ama}, holographically dual to  single entangled Bell pairs in the field theory, provide a natural and interesting environment where these ideas may be investigated. Indeed, the worldsheet theory of a  string may be viewed as the simplest theory of `quantum gravity' that one can solve \cite{Dubovsky:2012wk}. Even though worldsheet theories do not contain explicit gravitational degrees of freedom, it is important to remember that they have several surprising qualities that are similar to gravity, including the absence of local off-shell observables, a minimal length, a maximum achievable Hagedorn temperature \cite{Dubovsky:2012wk}. Further, their spectrum includes (integrable relatives of) black holes, which display fast scrambling and maximal chaos \cite{Murata:2017rbp,deBoer:2017xdk}. This is evident from the fact that their low energy effective action includes a soft sector governed by a Schwarzian action which couples with other modes in the infrared \cite{Banerjee:2018twd,Banerjee:2018kwy,Vegh:2019any}, reminiscent of two-dimensional dilaton theories and the SYK model. Finally, in some cases worldsheet theories may be cast as $T\bar{T}$-deformed field theories \cite{Cavaglia:2016oda,Callebaut:2019omt,Chakraborty:2019mdf}, which have proved crucial for studies of holography at a finite cutoff, e.g., \cite{McGough:2016lol,Dubovsky:2017cnj,Kraus:2018xrn}.

This paper is organized as follows. We begin in Section \ref{sec:prelims} with a quick review of open strings on AdS and the holographic dictionary for the worldsheet theory. Focusing on string states with an induced two-sided AdS$_2$ black hole geometry, we then discuss the computation of correlation functions using the extrapolate dictionary and sketch the derivation of two- and four-point functions, appearing in previous literature. In Section \ref{sec:travw} we consider the effects of a deformation that may naturally arise from interactions between the endpoints of the string. At the lowest order, this coupling is of the form $\delta\mathcal{L}\sim h\partial\phi_L\partial\phi_R$, in contrast with the standard deformation $\delta\mathcal{L}\sim h\phi_L\phi_R$ studied by Gao, Jafferis and Wall \cite{Gao:2016bin}. The effects of this deformation are studied in detail from the worldsheet perspective, drawing a comparison to the standard case that was considered in earlier research on AdS$_2$ holography \cite{Maldacena:2017axo}.\footnote{It is also tempting to speculate about the boundary dual of such deformation. Qualitatively, we expect it should act similarly to a deformation of the type $\delta\mathcal{L}\sim h T_L\bar{T}_R$, recently considered in \cite{Ferko:2022dpg}.} The string coupling is irrelevant in the dual CFT$_1$ and, as expected from scaling arguments, does not lead to a large effect in the IR other than slightly opening the wormhole throat. However, the coupling induces a large backreaction in the UV, effectively pulling the AdS boundaries inward. This effect ultimately improves the bound on information transfer between one boundary to another, effectively rendering the teleportation protocol more efficient. We close in Section \ref{sec:discussion} with a discussion of our results and open questions that remain for future work. 

\section{Preliminaries: open strings and AdS$_{2}$/CFT$_{1}$\label{sec:prelims}}

\subsection{Classical string solutions and worldsheet black holes}

Let us start by describing our holographic setup. We will consider a generic bulk geometry of the form $\text{AdS}_{d+1}\times X^p$, dual to some $d$-dimensional large-$N_c$ field theory. We then consider a stack of $N_{f}$ flavor branes on top of this geometry, with $N_{f}\ll N_{c}$. In this limit, the flavor branes act as probe branes and one may neglect their backreaction. The flavor branes will be assumed to span all directions of the dual theory, however, they will have compact radial support, spanning from boundary $z=0$ up to some cutoff surface $z_{c}$, where they end smoothly (meaning that one of their internal cycles shrinks down to zero size). A prototypical example is the standard D3/D7 system \cite{Karch:2002sh}, however, will be largely agnostic about the precise details of the top-down construction.

We then consider a fundamental open string in such AdS space, with its two endpoints attached to one of the flavor branes.\footnote{The string is dual to a pair of infinitely heavy quarks in the limit $z_c\to0$. Finite $z_c$ introduces a scale that can be associated with the mass of the quarks \cite{Karch:2002sh}.}
The dynamics of the string are governed by the standard Nambu-Goto action,
\begin{equation}\label{NGAction}
S_{\text{NG}} = -\frac{1}{2\pi\alpha'} \int_{\Sigma} \sqrt{-\text{det}\,\gamma_{\alpha\beta}}\,,
\end{equation}
where $\gamma_{\alpha\beta}=g_{\mu\nu}\partial_\alpha X^\mu\partial_\beta X^\nu$ is the induced metric on the worldsheet and $X^{\mu}{(\tau,\sigma)}$ are the embedding functions into the target space.

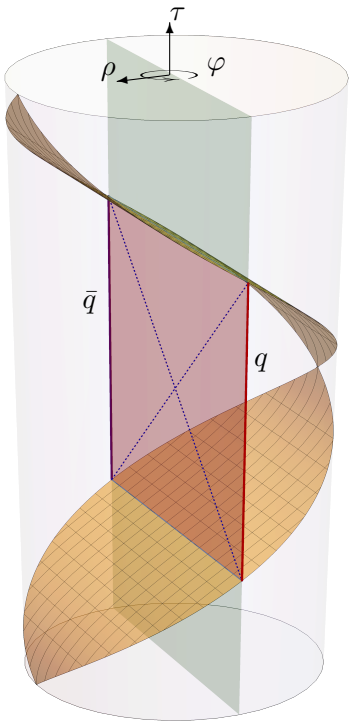
\begin{figure}[t!]
\centering
\includegraphics[width=2.5in]{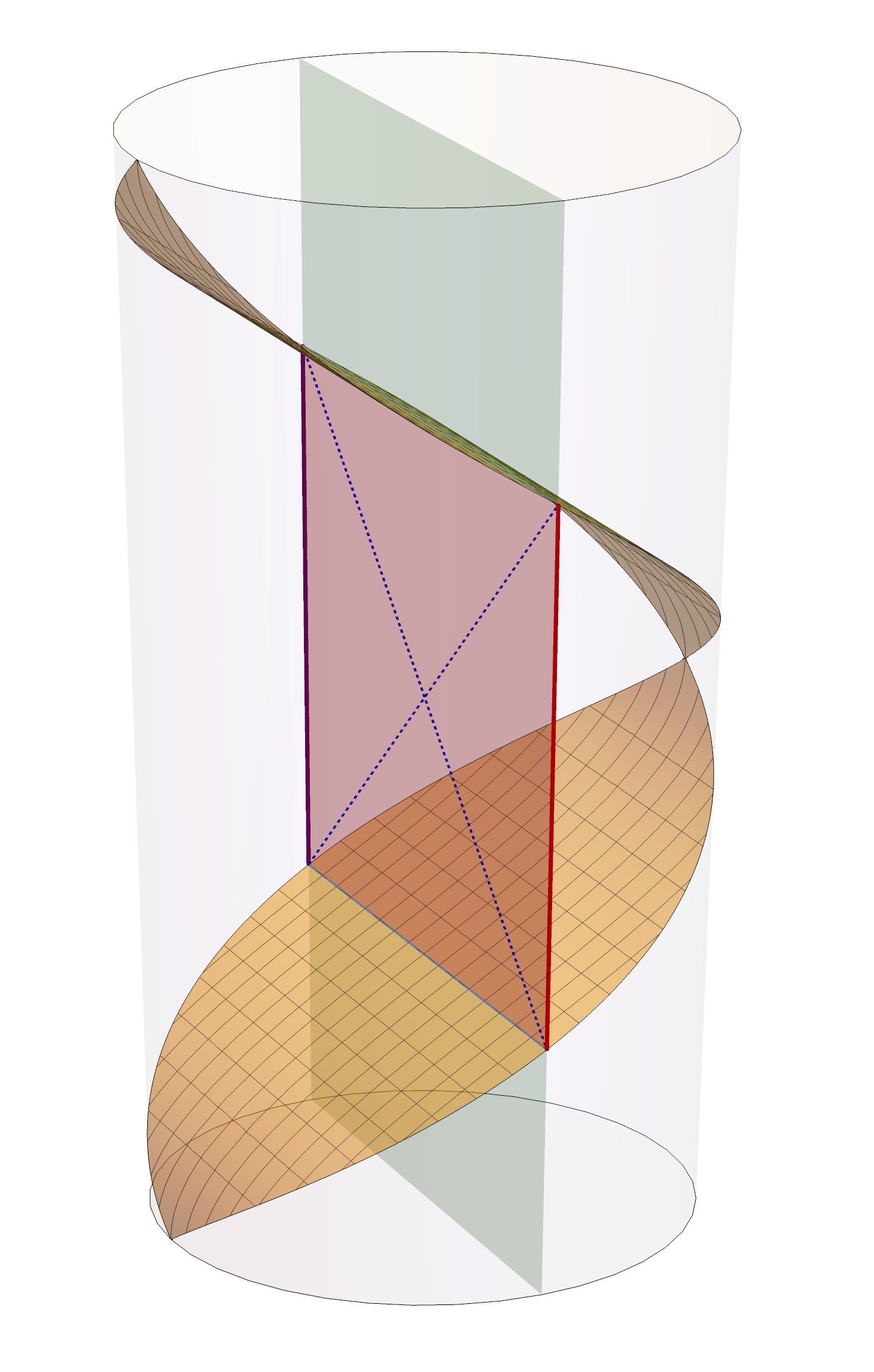}
\put(-128,170){$\bar{q}$}
\put(-63,148){$q$}
\put(-95,258){\vector(0,1){20}}
\put(-115,258){\rotatebox{97}{\vector(0,1){20}}}
\put(-95,279){$\tau$}
\put(-121,258){$\rho$}
\put(-111,260){\scalebox{3}[.5]{\rotatebox{200}{$ \circlearrowleft$}}}
\put(-81,260){$\varphi$}
 \caption{Static string dual to a quark $q$ and anti-quark $\bar{q}$ pair in global AdS. The two null planes depicted in yellow delimit the interior of a Poincar\'e patch, as seen by an accelerated observer. From the point of view of this Poincar\'e observer, the $q\bar{q}$-pair undergoes back-to-back constant acceleration. The induced worldsheet geometry hence includes a two-sided horizon and a wormhole, which is classically non-traversable. \label{figGlobal}}
\end{figure}

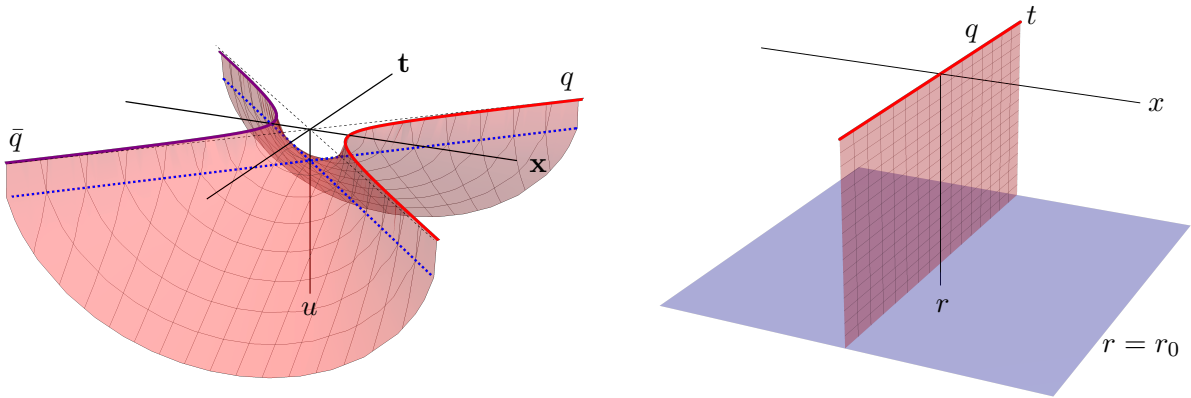
\begin{figure}[t!]
\centering
\includegraphics[width=3.2in,trim=1in 0 0 0]{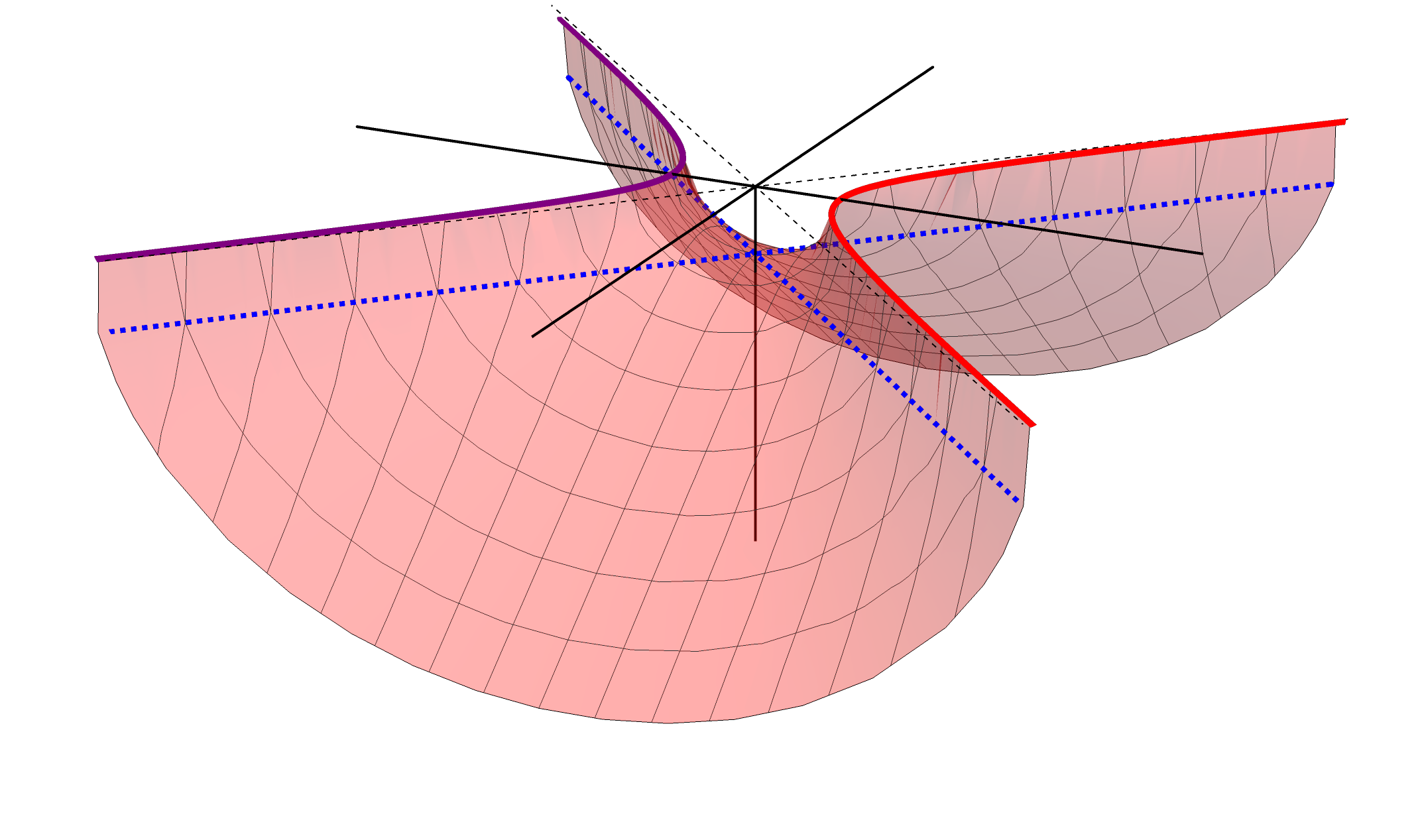}\includegraphics[width=3.2in,trim=0.3in 0 0 0]{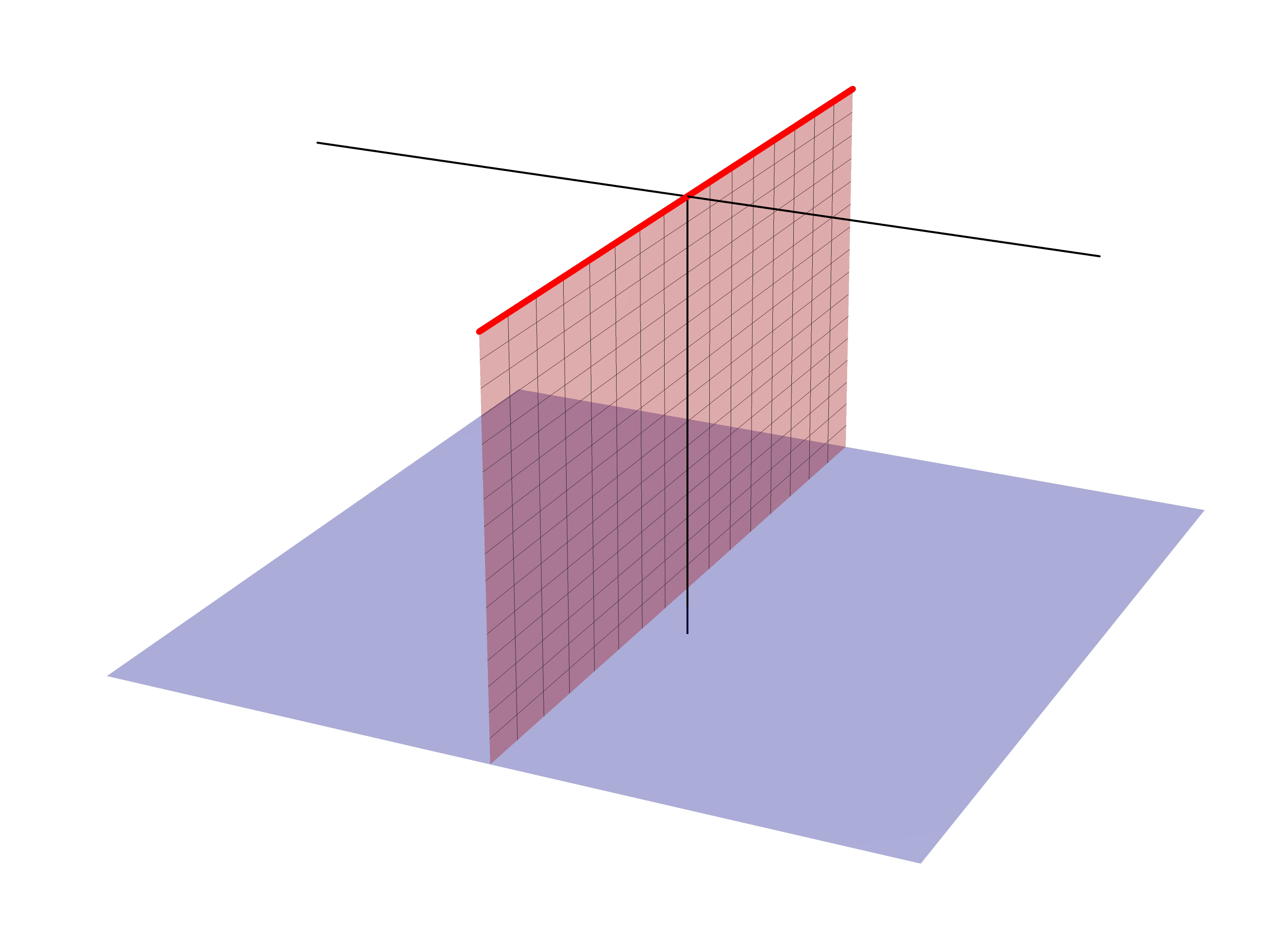}
 \put(-315,136){$\mathbf{t}$}
 \put(-355,44){ $u$}
 \put(-265,98){$\mathbf{x}$}
 \put(-257,131){ $q$}
 \put(-462,108){$\bar{q}$}
  \put(-77,154){$t$}
 \put(-115,45){ $r$}
 \put(-31,121){$x$}
 \put(-104,149){ $q$}
 \put(-52,30){ $r=r_0$}

 \caption{Left: uniformly-accelerated quark-anti-quark pair ($q\bar{q}$) and their corresponding string dual, plotted in the Poincar\'e patch of AdS. The worldsheet geometry includes a two-sided horizon and a wormhole, which is classically non-traversable. Right: string embedding in a Rindler patch of AdS. This coordinate system only covers one side of the solution, corresponding to one of the quarks. The solution is static and stretches between the boundary of AdS and an accelerating horizon that develops in the bulk. \label{figPoincare}}
\end{figure}

We will now consider global AdS$_3$ for concreteness, but the generalization to higher dimensions is straighforard. The bulk metric reads 
\begin{align}
ds^2=-\left(1+\frac{\rho^2}{\ell^2}\right)d\tau^2+\rho^2d\varphi^2+{d\rho^2\over \left(1+\frac{\rho^2}{\ell^2}\right)}\,.
\end{align}
It can be checked that a static string lying at a fixed azimuthal angle,
\be
X^{\mu}(\tau,\rho)=\{\tau,\varphi(\tau,\rho)= \text{constant},\rho\}\,,
\ee
solves the equations of motion and is therefore a valid solution. This situation is depicted in Fig. \ref{figGlobal}.
One may wonder how this solution looks like from the point of view of a Poincar\'e observer, which only has access to part of the spacetime ---see Fig. \ref{figGlobal} for an illustration. The bulk metric in this case takes the form
\begin{align}
ds^2=\frac{\ell^2}{u^2}\left[-d\mathbf{t}^2+d\mathbf{x}^2+du^2\right]\,,
\end{align}
and the string profile maps to the well-known solution with constant acceleration \cite{Hubeny:2014kma,Xiao:2008nr},
\be
X^{\mu}(\mathbf{t},u)=\{\mathbf{t},\mathbf{x}(\mathbf{t},u)=\sqrt{A^{-2}+\mathbf{t}^2-u^2},u\}\,.
\ee
This profile is shown in Fig. \ref{figPoincare}.
The accelerated solution contains a two-sided horizon and a wormhole in the induced worldsheet geometry \cite{Chernicoff:2013iga}. As pointed out in this paper, one may deform this solution in a number of ways (as allowed by the classical equations of motion) but always in a way such that the corresponding wormhole remains \emph{non-traversable}.

One can further specialize to the case of a Rindler observer in the bulk, in which case, only part of the string embedding is accessible. The bulk metric in this case takes the form of a (planar) BTZ black brane
\begin{align}
ds^2=-r^2f(r)d t^2+r^2dx^2+{d r^2\over r^2f(r)},\qquad f(r)=1-\left({r_0\over r}\right)^{2}\,,
\end{align}
and the embedding corresponds to a static vertical string that stretches between the AdS boundary and the horizon \cite{Caceres:2010rm}
\be
X^{\mu}(t,r)=\{t,x(t,r)=0,r\}\,.
\ee
The induced metric on the worldsheet takes the form of a 2d black hole:
\be\label{inducedg}
\gamma_{\alpha\beta}=\left(
                       \begin{array}{cc}
                         -r^2f(r) & 0 \\
                         0 & \frac{1}{r^2f(r)} \\
                       \end{array}
                     \right),
\ee
which coincides with a constant-$x$ section of a BTZ black brane. Perturbations over the static string embedding thus correspond to perturbations on top of this 2d black hole. Since black holes are known to be fast scramblers and saturate the bound on the quantum Lyapunov exponent $\lambda_L$, this lead to the conjecture that open strings could also exhibit maximal chaos. Indeed, explicit calculation of the four-point OTOC \cite{deBoer:2017xdk} confirmed this intuition.

\subsection{Semi-classical worldsheet theory}

The worldsheet theory for fluctuations around the static string embedding can be understood as a QFT living on a black hole background. To see the details of this theory, we can expand the NG action around this solution, i.e., $X^\mu=X_0^\mu+\delta X^\mu$, where
\be\label{flucts}
X_0^\mu(t,r)=\{t,x(t,r)=0,r\}\,,\qquad \delta X^\mu(t,r)=\{0,\delta x(t,r)\equiv\phi(t,r),0\}\,.
\ee
Plugging this into the action we obtain:
\be \label{eq-action}
S=S_{\text{free}}+S_{\text{int}}\,,
\ee
where $S_{\text{free}}$ is the action of the free theory and $S_{\text{int}}$ is the interaction piece, respectively:
\be\label{termsS}
S_{\text{free}}=\frac{1}{2}\int dt dr\left(\frac{\dot{ \phi}^2 }{f}-r^4f \phi'^2\right)\,,\qquad S_{\text{int}}=\frac{\kappa}{4}\int dt dr\left(\frac{ \dot{ \phi }^2 }{f}-r^4f \phi'^2\right)^2+\cdots\,.
\ee
For convenience, we have rescaled $\phi \to\sqrt{2\pi\alpha'}\phi$ and set $\kappa=\pi\alpha'$. Also, notice that we have only included interactions up to quartic order, which are relevant for the calculation of two- and four-point functions at leading order in $\alpha'$. In the next subsection, we will review the calculation of these correlators using the well-known `extrapolate' prescription.

\subsubsection{Two- and four-point functions}
The equation of motion that follows from $S_{\text{free}}$ is:
\be\label{waveeq}
[-f^{-1}\partial_t^2+\partial_r(r^4f\partial_r )]\phi(t,r)=0\qquad\rightarrow\qquad [f^{-1}\omega^2+\partial_r(r^4f\partial_r )]g(r)=0\,,
\ee
where we have set $\phi (t,r)=e^{-i\omega t}g(r)$. The two linearly independent solutions to this equation can be written as $g_{\pm \omega}(r)$ where:
\be\label{slns}
g_{\omega}(r)={1\over 1+ i\nu}{\rho+ i\nu \over \rho}\left({\rho-1\over \rho+1}\right)^{i\nu/2}\,,
\ee
and we have defined the following dimensionless variables:
\be
 \rho\equiv {r\over r_0},\qquad
 \nu\equiv{\omega\over r_0}={\beta \omega\over 2\pi}\,.
\ee
These solutions satisfy $g_\omega(r)^*=g_{-\omega}(r)$ and are purely outgoing/ingoing, respectively. To see this, we can define the
tortoise coordinate $r_*$ such that
\be\label{tort}
dr_*^2=\frac{dr^2}{r^4f(r)^2}\,.
\ee
This coordinate is convenient because the worldsheet metric becomes conformally flat and the fluctuations are easier to analyze.
Integrating (\ref{tort}) we find
\be\label{tort2}
r_*=-\frac{1}{r_0}\text{arccoth}\left(\frac{r}{r_0}\right)\,.
\ee
As we can see, the horizon is mapped to $r_*\to-\infty$ while the boundary is now at $r_*\to0$. In particular, in the near-horizon region, we have
\be
 r_*\sim {1\over 2r_0}\log\left({r\over r_0}-1\right)\,,
\ee
and the equation of motion becomes
\be
(\omega^2+\partial_{r_*}^2)g(r)\approx0\,.
\ee
It is easy to see that the solutions should take the following form:
\be
g(r)\approx e^{i\omega r_*}\quad\text{(outgoing)}\,,\qquad g(r)\approx e^{-i\omega r_*}\quad \text{(ingoing)}\,.
\ee
In terms of the tortoise coordinate, (\ref{slns}) becomes:
\be
 g_{\omega}(r) = {1\over 1+ i\nu}{\rho+ i\nu \over \rho}\, e^{i\omega r_*}\,,
\ee
so we can indeed identify $g_{\pm\omega}(r)$ as outgoing/ingoing solutions. Next, we impose a Neumann boundary condition for $\phi(r,t)$ in the UV,\footnote{Normally, one would choose normalizable boundary conditions in the UV. However, that would correspond to a string extending all the way to the AdS boundary. The mass of the dual particles would be infinite and the correlators would trivially vanish. Instead, we introduce a UV cutoff $r_c$ to make the masses finite and allow for string fluctuations in the UV. We implement this by means of a Neumann boundary condition at $r=r_c$.}
\be
\partial_r \phi(r_c,t)=0\,,
\ee
where $r_c=1/z_c$ is a UV cutoff.
This condition dictates that we take the linear combination:
\be
 f_\omega(r)=g_\omega(r)+e^{i\theta_\omega}g_{-\omega}(r)\,,\qquad e^{i\theta_\omega}=-\frac{\partial_r g_\omega(r_c)}{\partial_r g_{-\omega}(r_c)}\,.\label{f_ito_g_app}
\ee
It is easy to see that the phase $\theta_\omega$ is real. The field $\phi$ can now be expanded as follows:
\begin{align}\label{Xexp}
 \phi(t,r)={\sqrt{2\pi\ap}\over r_H}\int_0^\infty {d\omega\over 2\pi}
 {1\over \sqrt{2\omega}}
 \left[ f_\omega(r)e^{-i\omega t}a_\omega
 + f_\omega(r)^* e^{i\omega t}a_\omega^\dagger\right],
\end{align}
The normalization constant in front of (\ref{Xexp}) appears after properly normalizing the mode functions $f_\omega(r)$ \cite{Atmaja:2010uu}. Moreover,
since the system is at finite temperature $T$, the expectation value of the creation and annihilation operators should follow a Bose distribution:
\be
\langle a^\dagger_{\omega}a_{\omega'}\rangle={2\pi\delta(\omega-\omega')\over e^{\beta\omega}-1}\,.
\ee

We will be interested in the Wightman and retarded propagators, defined as follows:
\begin{align}\label{defprop}
\begin{split}
  D_W(t-t',r,r')    &=\ev{\phi(t,r)\phi(t',r')}\,,\\
 D_{\rm ret}(t-t',r,r')&=\theta(t-t')\ev{[\phi(t,r),\phi(t',r')]}\,.
\end{split}
\end{align}
Using the wave
equation (\ref{waveeq}) and the canonical commutation relation
\be
 [\phi(t,r),\dot{\phi}(t,r')]={2\pi i\ap }f \delta(r-r')\,,
\ee
it can be shown that these propagators satisfy
\begin{align}
[-f^{-1}\partial_t^2+\partial_r(r^4f\partial_r )]
 D_{W}(t-t',r,r')
 &=0\,,\\
[-f^{-1}\partial_t^2+\partial_r(r^4f\partial_r )]
 D_{\text{ret}}(t-t',r,r')
 &={2\pi i \ap}\delta(t-t')\delta(r-r')\,.\label{waveeq_D_F,ret,adv}
\end{align}
It is then easy to show that the Wightman
propagator can be written as
\begin{align}
 D_W(\omega,r,r')&=
 {2\pi\ap\over r_0^2}
 {f_{\omega}(r)f_{-\omega}(r')\over 2\omega(1-e^{-\beta\omega})},\label{D_W_ito_f}
\end{align}
where $f_{-\omega}=f_{\omega}^*$. These correlators can also be derived using the Schwinger-Keldysh path integral formalism \cite{Banerjee:2013rca}.

Depending on time ordering, there are a few possible four-point correlation functions that will be of interest to our purposes. These can be written as
\be \label{eq-otoc1}
D(t_1,t_2,t_3,t_4)= \langle W(t_1) V(t_2) W(t_3) V(t_4)\rangle\,.
\ee
In this case, one may consider the standard Schwinger-Keldysh contour or a more general time-fold to obtain exotic time orderings (see e.g. \cite{Shenker:2014cwa}). 
The relevant one to study chaos is the out-of-time-order (OTOC) for which $t_1>t_2$ and $t_3>t_4$ but $t_2<t_3$ and $t_4<t_1$. This correlator was studied in \cite{deBoer:2017xdk} for the case at hand. The OTOC calculation boils down to computing a scattering amplitude in the near-horizon region, which may be obtained using the eikonal approximation. This gives the following result \cite{Shenker:2014cwa}
\be
D_\text{otoc}(t_1,t_2,t_3,t_4)= \int_{0}^{\infty} dp\, \psi_1(p) \psi^{*}_3(p) \int_{0}^{\infty} dq \,  \psi_2(q) \psi_4^{*}(q) \,e^{i \delta}\,,
\ee
where the wave functions $\psi_i$ are Fourier transforms of bulk-to-boundary correlators
\bea
\psi_1(p) &=&\int dv e^{2 i p v} \langle \varphi_V(u,v) V(t_1)^{\dagger} \rangle\big|_{u=0}\,,\\
\psi_2(q) &=&\int du e^{2 i p u} \langle \varphi_W(u,v) V(t_2)^{\dagger} \rangle\big|_{v=0}\,,\\
\psi_3(p) &=&\int dv e^{2 i p v} \langle \varphi_V(u,v) V(t_3) \rangle\big|_{u=0}\,,\\
\psi_4(q) &=&\int du e^{2 i p u} \langle \varphi_W(u,v) V(t_4) \rangle\big|_{v=0}\,,
\eea
$\varphi_V$ and $\varphi_W$ are the worldsheet fields dual to the boundary operators $V$ and $W$, and $u$ and $v$ are Kruskal coordinates (see (\ref{eq-kruskal})). The phase shift is given by
\be
\delta(s) = \frac{s \ell_s^2}{4}\,,
\ee
where $s=(\Delta E)^2$ is the standard Mandelstam variable and $\ell_s\equiv\sqrt{2\pi\alpha'}$. Evaluating (\ref{eq-otoc1}) at leading order in $\ell_s^2$ for $t_1= i \epsilon_1, t_2= t+ i \epsilon_2, t_3= i \epsilon_3$ and $t_4=t+ i \epsilon_4$, and normalizing the correlator with respect to the leading order term, one finds \cite{deBoer:2017xdk}
\be \label{eq-otoc2}
\frac{\langle V(i\epsilon_1) W(t+i\epsilon_2) V(i\epsilon_3) W(t+i\epsilon_4) \rangle }{\langle V(i\epsilon_1)V(i \epsilon_3)\rangle  \langle W(t+i\epsilon_2)W(t+i \epsilon_4)  \rangle}=1+\frac{2\, i \, \ell_s^2}{\epsilon_{13} \epsilon_{24}^*} e^{2\pi T t}\,,
\ee
where $\epsilon_{mn}=i \left( e^{i 2\pi T \epsilon_m} -e^{i 2\pi T \epsilon_n}\right)$. From (\ref{eq-otoc2}) we can read off the Lyapunov exponent $\lambda_L= 2\pi T$, which saturates the chaos bound \cite{Maldacena:2015waa}\,.

\section{Making worldsheet traversable wormholes\label{sec:travw}}

\subsection{Introducing a double trace deformation}

Following the seminal work of Gao, Jafferis and Wall \cite{Gao:2016bin}, one may wonder if we can consider the effects of a non-local coupling to construct traversable wormholes in the worldsheet. In order to do so, we must first ask what the natural form of such a coupling may be. To start with, we note that each endpoint of the string can be coupled to a flavor brane via the boundary term (see e.g. \cite{Chernicoff:2009ff})
\be\label{coupling}
S_F=\int_{\partial\Sigma} A_{\mu} \dot{ X }^\mu \,.
\ee
Normally, $A_{\mu}$ is meant to be taken as an external field on the brane, which one can turn on by hand. However, in our context, it is natural to take inspiration from the form of the backreacted field that arises from the presence of the second endpoint of the string. For a point-like source, this field is expected to follow the standard Li\'enard-Wiechert form, $A_{\mu}\sim \delta\dot{X}_\mu$. In our conventions (\ref{flucts}), then, we are then led to consider a coupling of the form:\footnote{The form of the potential $A_{\mu}\sim \delta\dot{X}_\mu$ could also be justified since this is the only choice that makes $S_F$ symmetric between the left and right boundaries/endpoints of the string.}
\be\label{nonlocal}
S_F= h \int_{\partial\Sigma}   \dot{ \phi }_L \dot{ \phi }_R \,,
\ee
where $h$ is a parameter that controls the strength of the backreaction.
This is a bilocal coupling, as in \cite{Gao:2016bin}, but involving derivatives. It is important to emphasize that backreaction itself is not enough to obtain such a coupling, as (\ref{nonlocal}) requires to properly account for retarded times and, hence, cannot possibly be instantaneous.  Nevertheless, terms like (\ref{nonlocal}) are quite natural to describe the leading interactions between endpoints since it is $\partial \phi$, and not $\phi$, that represents a primary operator from the worldsheet perspective. In the following, we will thus proceed by assuming that such a coupling has been turned on and study its implications.

To construct the traversable wormhole it would suffice to consider the free theory $S_{\text{free}}$ (free massless scalar field) together with the non-local coupling (\ref{nonlocal}), i.e., one can ignore the higher order interactions in a first approximation. However, the interaction terms in $S_{\text{\text{int}}}$ are important in order to actually study the information transfer, as one must take them into account for the computation of four- and higher-point correlators.

One difference between our work and the standard setup considered by Gao, Jafferis and Wall \cite{Gao:2016bin} is that our worldsheet metric $\gamma_{\alpha\beta}$ does not satisfy the semiclassical Einstein equation,
\be
R_{\alpha\beta}-\frac{1}{2}R \gamma_{\alpha\beta}+\Lambda \gamma_{\alpha_\beta}=\langle T_{\alpha\beta}\rangle\,,
\ee
or its equivalent in 2d dilaton gravity. Instead, it satisfies the (perhaps simpler) semiclassical equation
\be \label{eq-gamma}
\gamma_{\alpha\beta}=g_{\mu\nu}\langle\partial_\alpha X^\mu\partial_\beta X^\nu\rangle\,.
\ee

\subsection{Condition for traversability} \label{sec-traversability}
\begin{figure}[t!]
\centering

\begin{tikzpicture}[scale=1.5]
\draw [thick]  (0,0) -- (0,3);
\draw [thick]  (3,0) -- (3,3);
\draw [thick,dashed]  (0,0) -- (3,3);
\draw [thick,dashed]  (0,3) -- (3,0);
\draw [thick,decorate,decoration={zigzag,segment length=1.5mm, amplitude=0.4mm}] (0,3) -- (3,3);
\draw [thick,decorate,decoration={zigzag,segment length=1.5mm,amplitude=.4mm}]  (0,0) -- (3,0);

\draw[thick,<->] (1,2.2) -- (1.5,1.7) -- (2,2.2);


\node[scale=0.8, align=center] at (1.5,2.65) {Future Interior};
\node[scale=0.8,align=center] at (1.5,.55) {Past Interior};
\node[scale=0.8,align=center] at (0.6,1.6) {Left\\ Exterior};
\node[scale=0.8,align=center] at (2.4,1.6) {Right\\ Exterior};
\end{tikzpicture}
\vspace{0.1cm}
\put(-142,50){\rotatebox{90}{\small $r = \infty$}}
\put(-0,80){\rotatebox{-90}{\small $r = \infty$}}
\put(-75,-8){\small $r = 0$}
\put(-75,135){\small $r = 0$}
\put(-96,95){\small $v$}
\put(-46,95){\small $u$}

\caption{ \small Penrose diagram for the eternal AdS black hole described by (\ref{eq-metricKruskal}). In our conventions, time increases as one goes up (down) along the right (left) asymptotic boundary.}
\label{fig-Penrose}
\end{figure}
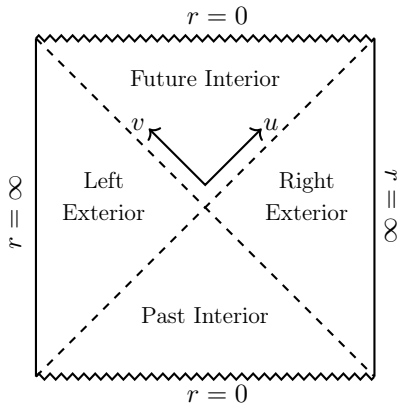

To understand the conditions under which the worldsheet wormhole becomes traversable, it is convenient to use Kruskal coordinates $(u,v)$:
\be \label{eq-kruskal}
e^{2 r_0 t}=-\frac{u}{v}\,,\,\,\,\, \frac{r}{r_0}=\frac{1-u v}{1+ u v}\,,
\ee
and parametrize the embedding as $X^{\mu}=\left( u,v,X(u,v) \right)$, which corresponds to a string stretching between the two asymptotic boundaries of an eternal AdS black hole. Considering a static solution $X(u,v)=0$, the worldsheet metric can be written as
\be \label{eq-metricKruskal}
ds_\text{ws}^2=-\frac{4 du dv}{(1+u v)^2}\,.
\ee
The corresponding Penrose diagram is shown in Fig. \ref{fig-Penrose}.

We note that a null ray originating from the left asymptotic boundary and moving along the $v=0$ horizon reaches the future singularity, never crossing to the right exterior region. One might wonder whether fluctuations of the geometry (\ref{eq-metricKruskal}) induced by string fluctuations can lead to a traversable wormhole. To analyze this, we start by parameterizing the string worldsheet with null coordinates $(u,v)$ and consider perturbations $\delta x(u,v)=\phi(u,v)$ around the static solution. 
In this case, the worldsheet metric (\ref{eq-metricKruskal}) becomes
\be
ds^2 = -\frac{4 du dv}{(1+u v)^2} + \bar{\gamma}_{\alpha \beta} dx^{\alpha} dx^{\beta}\,,
\ee
where
\be\label{flucmetric}
\bar{\gamma}_{\alpha \beta}=g_{x x} \langle\partial_\alpha \phi \partial_\beta \phi \rangle\,.
\ee
Next, we consider a perturbation moving in the $v$ direction and take $\phi(u,v)=F(u)$, where $F(u)$ is assumed to vanish outside a window around $u=u_0$. It is easy to check that only $\bar{\gamma}_{uu}$ is non-zero, and is given by
\be
\bar{\gamma}_{uu} = \left( \frac{1- u v}{1+ u v} \right)^2 F'(u)^2\,.
\ee
We can now determine how the above perturbation changes the trajectory of a null ray originating from the left asymptotic boundary and moving along the $v=0$ horizon. Let $u$ be the affine parameter along this ray. Along the trajectory of the ray, we have
\be
\left[ -\frac{4 du dv}{(1+u v)^2}+\left( \frac{1- u v}{1+ u v} \right)^2 F'(u)^2 du^2 \right]_{v=0} =0\,,
\ee
which implies
\be \label{eq-diffshift}
dv =\frac{F'(u)^2}{4} du\,.
\ee
By integrating (\ref{eq-diffshift}), we find that the null shift that the ray suffers as it crosses the perturbation at $u=u_0$ is given by
\be
\Delta v (u) = \frac{1}{4} \int_{-\infty}^{u} F'(u)^2 du\,.
\ee
Note that $\Delta v$ may be related to the energy of the perturbation, which is given by
\cite{deBoer:2017xdk}
\be
\Delta E = 2 \int du F'(u)^2\,.
\ee
Therefore
\be
\Delta v  =\frac{\Delta E }{8}\,.
\ee
The worldsheet wormhole becomes traversable if $\Delta v <0$, which implies $\Delta E<0$, i.e., the traversability requires the presence of a perturbation with negative energy. Of course, negative energy is forbidden for classical perturbations and so are worldsheet traversable wormholes \cite{Chernicoff:2013iga}. This condition is analogous to the so-called Average Null Energy Condition (ANEC), which needs to be overcome for wormhole traversability to be possible. In the next subsection, we will see that quantum effects in the worldsheet, together with the double trace coupling (\ref{nonlocal}), may indeed lead to a configuration where traversability is allowed.

\begin{figure}[t!]
\centering

\begin{tikzpicture}[scale=1.5]
\draw [thick]  (0,0) -- (0,3);
\draw [thick]  (3,0) -- (3,3);
\draw [thick,dashed]  (0,0) -- (3,3);
\draw [thick,dashed]  (0,3) -- (3,0);

\draw [thick,blue,decorate,decoration={snake,segment length=3mm,amplitude=0.5mm}]  (3,1.5) -- (1.5,2.9);
\draw [thick,blue,decorate,decoration={snake,segment length=3mm,amplitude=0.5mm}]  (3.03,1.53) -- (1.53,2.93);

\draw [thick,blue,decorate,decoration={snake,segment length=3mm,amplitude=0.5mm}]  (0,1.5) -- (1.5,2.9);
\draw [thick,blue,decorate,decoration={snake,segment length=3mm,amplitude=0.5mm}]  (-0.03,1.53) -- (1.47,2.93);

\draw [red,thick,->]  (0,0.2) -- (1.54,1.74);
\draw [red,thick]  (1.54,1.74) -- (2.14,2.34);
\draw [red,thick,->]  (2.34,2.14) -- (2.7,2.5);
\draw [red,thick]  (2.7,2.5) -- (3,2.8);

\draw [thick,decorate,decoration={zigzag,segment length=1.5mm, amplitude=0.3mm}] (0,3) -- (3,3);
\draw [thick,decorate,decoration={zigzag,segment length=1.5mm,amplitude=.3mm}]  (0,0) -- (3,0);

\draw[thick,<->] (-0.4,3.4) -- (-0.1,3.1) -- (0.2,3.4);

\end{tikzpicture}
\vspace{0.1cm}
\put(-150,60){\rotatebox{0}{\small $V_L$}}
\put(-1,60){\rotatebox{0}{\small $V_R$}}
\put(-170,10){\rotatebox{0}{\small $W_L(t_1)$}}
\put(-1,120){\rotatebox{0}{\small $W_R(t_2)$}}
\put(-157,150){\rotatebox{0}{\small $v$}}
\put(-125,150){\rotatebox{0}{\small $u$}}

\caption{\small The non-local coupling between $V_L$ and $V_R$ introduces a negative-energy shock wave in the bulk that makes the wormhole traversable. The traversability can be diagnosed by a two-sided correlation function involving $W_L$ and $W_R$.}
\label{fig-info-tranfer}
\end{figure}
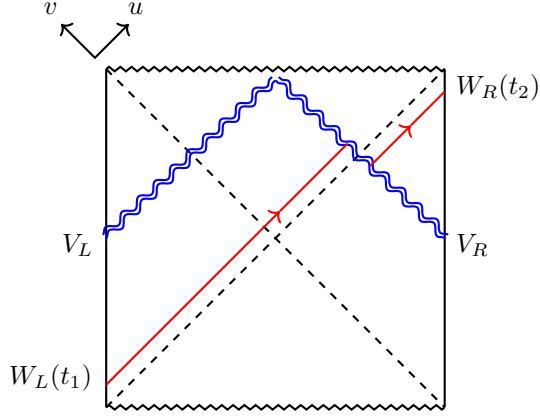

\subsection{Wormhole opening}

Inspired by  \cite{Gao:2016bin}, we now compute the opening of the wormhole using a point-splitting technique. We start with the semiclassical equation
\be \label{eq-gamma}
\gamma_{\alpha_\beta}=g_{\mu\nu}\langle\partial_\alpha X^{\mu} \partial_\beta  X^{\nu} \rangle\,,
\ee
and its fluctuation version (\ref{flucmetric}), from which we can obtain the opening of the wormhole as
\be\label{wormop}
\Delta v (u) = -\frac{1}{2 \gamma_{uv}(0)} \int_{-\infty}^{u} \bar{\gamma}_{uu} du\,.
\ee
In the point splitting method, we can compute the metric perturbation (\ref{eq-metricKruskal}) as
\be \label{eq-pointsplitting}
 \bar{\gamma}_{\alpha \beta}   = \lim_{\sigma' \rightarrow \sigma} \partial_{\alpha} \partial'_{\beta} G(\sigma,\sigma')\,,
\ee
where $\sigma =(u,v)$ and $\sigma'=(u',v')$ denote worldsheet points, and $G(\sigma, \sigma')$ is a renormalized two-point function under the presence of a Hamiltonian deformation $\delta H(t)$:
\bea \label{eq-Gh3}
G(u,v;u',v')&=& g_{xx} \, \langle  \phi_H(u,v)  \phi_H(u',v')  \rangle=g_{xx} \, \langle  \phi_H(t,r)  \phi_H(t',r')  \rangle\,, \nonumber \\
&=&g_{xx} \,  \langle U^{-1}(t,t_0) \phi_I(t,r) U(t,t_0) U^{-1}(t',t_0) \phi_I(t',r') U(t',t_0) \rangle\,,
\eea
where $U(t,t_0)= \mathcal{T} e^{-i \int_{t_0}^{t} dt \delta H(t)}$ denotes the evolution operator in the interaction picture, while $g_{xx}=\frac{(1-u v)^2}{(1+u v)^2}$. The subscripts $H$ and $I$ in the first and second lines of (\ref{eq-Gh3}) indicate a field in the Heisenberg and interaction pictures, respectively.  In what follows, we will suppress the subscript $I$. Further, we will consider fields in the right exterior region of the geometry.

We are interested in the $uu$ component of the worldsheet metric, which only involves derivatives of $G(u,v;u',v')$ with respect to $u$ and $u'$. This leads to the following simplification
\be
\bar{\gamma}_{u u} = \lim_{(u',v') \rightarrow (u,v)} \partial_u \partial_{u'} G(u,v;u',v') =  \lim_{u' \rightarrow u} \partial_u \partial_{u'} G(u,v;u',v)\,.
\ee
Moreover, we only focus on $\bar{\gamma}_{uu}$ on the horizon $v=0$, which leads to
\be
\bar{\gamma}_{uu} = \lim_{u' \rightarrow u} \partial_u \partial_{u'} G(u,0;u',0)\,.
\ee
Therefore, we only need to care about $G(u,0;u',0)$.  Note that, by taking $v=v'=0$ in (\ref{eq-Gh3}) we obtain
\be
G(u,0;u',0)= \langle U^{-1} \phi(u,0) U U^{-1} \phi(u',0) U \rangle\,.
\ee
We now compute $G(u,0;u',0)$ at first order in perturbation theory. We perform an expansion for small $h$ and write the two-point function $G(\sigma,\sigma')$ as
\be
G(\sigma, \sigma')=G_0(\sigma, \sigma')+G_h(\sigma, \sigma')\, h+ \cdots\,,\\
\ee
where $G_0$ is the unperturbed two-point function, while $G_h$ corresponds to the first correction induced by the deformation $\delta H$. Using (\ref{eq-Gh3}), we can see that
\be
G_h = - i \int_{t_0}^ {t} dt_1 \langle [\phi(t,r),\delta H(t_1)] \phi(t',r')\rangle-i \int_{t_0}^ {t'} dt_1 \langle \phi(t,r) [\phi(t',r'),\delta H(t_1)] \rangle\,.
\ee
Now we need to specify the deformation $\delta H$. We are going to consider two cases:

\begin{itemize}
\item Case I:
\be
\delta H(t) = h(t)\, \phi_L(-t) \phi_R(t) \,\,\,\, \text{with}\,\,\,\,h(t) = h \left( \frac{2\pi}{\beta} \right)^{1-2\Delta} \theta \left(\frac{2\pi}{\beta}(t-t_0) \right)\,,
\ee
\end{itemize}
and
\begin{itemize}
\item  Case II:
\be
\delta H(t) = h(t)\, \dot{\phi}_L(-t) \dot{\phi}_R(t) \,\,\,\, \text{with}\,\,\,\,h(t) = h \left( \frac{2\pi}{\beta} \right)^{-1-2\Delta} \theta \left(\frac{2\pi}{\beta}(t-t_0) \right)\,,
\ee
\end{itemize}
where $\beta=\frac{2\pi\ell^2}{r_0}$ is the inverse temperature. The scaling factors $( \tfrac{2\pi}{\beta})^{1-2\Delta}$ and $( \tfrac{2\pi}{\beta})^{-1-2\Delta}$ are introduced here for convenience, to make $h$ dimensionless. 

In the remaining of this section, we will set $r_0=\ell=1$ to lighten the notation, but they could be easily restored. The results for general values of $r_0$ can be found in Sec.~\ref{sec-commutator}. Further, we point out that the effects of the coupling I have been studied and reported in previous literature, originally in \cite{Gao:2016bin} for $d=2$ and in \cite{Freivogel:2019whb} for arbitrary $d$ (our case corresponds to $d=1$). We include the analysis of this coupling here for completeness.

\subsection*{Case I: $\delta H(t) = h(t)\, \phi_L(-t) \phi_R(t)$ }
In this case, we obtain\footnote{We only need to consider the contribution from $G_h$ because the contribution coming from $G_0$ is fixed by symmetry to be  proportional to the background metric. Therefore, $G_0$ just leads to a rescaling of the AdS length scale, which does not contribute to the wormhole opening $\Delta v$. Also, note $G_h(\sigma,\sigma')$ is finite even for $\sigma=\sigma'$, so the point-splitting is not essential to compute this contribution.}
\bea
i G_h &=&   \int_{t_0}^ {t} dt_1 h(t_1) \langle [\phi(t,r), \phi_L(-t_1) \phi_R(t_1)] \phi(t',r')\rangle+ \nonumber \\&&+  \int_{t_0}^ {t'} dt_1 h(t_1) \langle \phi(t,r) [\phi(t',r'), \phi_L(-t_1) \phi_R(t_1)] \rangle\,.
\eea
Factorizing the above four-point functions in terms of a product of two-point functions and using causality, we can write
\be
G_h = - i \int_{t_0}^ {t} dt_1 \, h(t_1)\langle \phi(t',r') \phi_L(-t_1)\rangle   \langle [\phi(t,r), \phi_R(t_1)]  \rangle + (t \leftrightarrow t')\,.
\ee
Note this expression is not invariant under $L\leftrightarrow R$ because there are two operator insertions on the left boundary. Next, we use the KMS condition~\cite{Haag:1967sg}:
\be
\langle \mathcal{O}_R(t) \mathcal{O}_L(t')\rangle_{\text{TFD}}=\langle \mathcal{O}_R(t) \mathcal{O}_R(t'+i\beta/2)\rangle_{\text{TFD}}\,,
\ee
and write the final result in terms of a product of bulk-to-boundary correlators
\be
G_h = 2 \sin (\pi \Delta) \int dt_1 h(t_1) K(t'+t_1-i \beta/2,r')K^r(t-t_1,r)+(t \leftrightarrow t')\,,
\ee
where
\bea
K(t-t',r) & \equiv & \langle \phi(t,r) \phi(t',\infty)\rangle\,,  \\
K^r(t-t',r) & =& |K(t-t',r)| \theta(t) \, \theta \left( \frac{\sqrt{r^2-r_0^2}}{r_0} \cosh(t-t')-\frac{r}{ r_0}  \right)\,,
\eea
denote the Wightman bulk-to-boundary propagator and the retarded bulk-to-boundary propagator, respectively. In AdS-Rindler coordinates $(t,r)$, the bulk-to-boundary propagator for a scalar field dual to an operator of dimension $\Delta$ is
\be
K_{\Delta}(t,r) = c_{\Delta} \left[ -\frac{\sqrt{r^2-r_0^2}}{r_0} \cosh(t-t')+\frac{r}{ r_0} \right]^{-\Delta} \,,
\ee
where
\be
c_{\Delta}=\frac{\Gamma(\Delta)}{2^{\Delta+1} \pi^{1/2} \Gamma(\Delta+1/2)}\,.
\ee
 For convenience, we will do the calculations for a generic scalar field of dimension $\Delta$ and we will set $\Delta=1$ at the end of our calculations. Furthermore, it is more convenient to use Kruskal coordinates $(u,v)$, in which terms the  propagator becomes
\be
K_{\Delta}(u,v;u_1) = c_{\Delta} \left[ \frac{1+uv}{v u_1-\frac{u}{u_1}+1-u v }\right]^{\Delta} \,.
\ee
Thus, evaluated at $v=0$ in Kruskal coordinates, $G_h$ becomes
\be \label{eq-Ghkruskal}
G_h(u,u')= 2 \ell_s^2 \sin(\pi \Delta) c_{\Delta}^2 \int\frac{d u_1}{u_1} h(u_1)  \frac{\theta \left( \frac{u}{u_1}-1 \right) }{(\frac{u}{u_1}-1)^{\Delta}} \frac{1}{(u_1 u'+1)^{\Delta}}+(u \leftrightarrow u') \equiv H(u,u')+H(u',u),
\ee
where $u_1=e^{r_0 t_1}$. Importantly, the above bulk-to-boundary propagators are appropriate for fluctuations with a canonical kinetic term, which is the case if we rescale the fluctuations as $\phi \rightarrow \ell_s \phi$ as in (\ref{termsS}). This rescaling introduces a factor of $\ell_s^2$ in $G_h$.

The $uu$ component of the perturbed worldsheet metric is then computed as
\be
\bar{ \gamma}_{uu} = 2\,h\, \lim_{u' \rightarrow u} \partial_u \partial_{u'} H(u,u')\,,
\ee
where $H(u,u')$ is defined in (\ref{eq-Ghkruskal}). Likewise, the wormhole opening (\ref{wormop}) now reads
\bea
\Delta v &=& -\frac{1}{2\gamma_{uv}(0)} \int_{u_0}^{u} \bar{\gamma}_{uu} du = \frac{1}{4}  \int_{u_0}^{u} du \,2 \lim_{u' \rightarrow u} \partial_u \partial_{u'} H(u,u')\,,\\
&=& \frac{1}{2} \int_{u_0}^{u} \lim_{u' \rightarrow u} \partial_u \tilde{H}(u,u';u_0)\,,
\eea
where
\be
 \tilde{H}(u,u';u_0) \equiv \partial_{u'} H(u,u') = C_0 \int_{u_0}^{u} du_1 \frac{u_1^{\Delta}}{(u-u_1)^{\Delta}(u' u_1+1)^{\Delta+1}}\,,
\ee
and with $C_0=2 h \Delta \sin(\pi \Delta) c_{\Delta}^2$.
Notice that
\be
\lim_{u' \rightarrow u} \partial_u  \tilde{H}(u,u';u_0)= \partial_u  \tilde{H}(u,u;u_0)-\partial_u^{(2)}  \tilde{H}(u,u;u_0)\,,
\ee
where $\partial_u^{(2)}$ denotes a derivative that acts only on the second argument of $ \tilde{H}(u,u;u_0)$. Using the above relation, we can write
\be
\int_{u_0}^{\infty} \lim_{u' \rightarrow u} \partial_u  \tilde{H}(u,u';u_0)= \tilde{H}(\infty, \infty; u_0)- \tilde{H}(u_0,u_0;u_0)-\int_{u_0}^{\infty} du \, \partial_u^{(2)}  \tilde{H}(u,u;u_0)\,.
\ee
Noticing that $ \tilde{H}(\infty, \infty; u_0)=0$ and $ \tilde{H}(u_0,u_0;u_0)=0$, we arrive at
\be
\Delta v = -\frac{C_0}{2} \int_{u_0}^{\infty} du\, \partial_u^{(2)}  \tilde{H}(u,u;u_0)=-\frac{C_0}{2} \int_{u_0}^{\infty} du \int_{u_0}^{u} du_1\, \frac{(\Delta+1) u_1^{\Delta+1}}{(u-u_1)^{\Delta}(u u_1+1)^{\Delta+2}}\,.
\ee
We now use the identity
\be
\int_{u_0}^{\infty} du \int_{u_0}^{u}du_1 \, \mathcal{F}(u,u_1)=\int_{u_0}^{\infty} du_1 \int_{u_1}^{\infty}du \, \mathcal{F}(u,u_1)\,,
\ee
with $\mathcal{F}(u,u_1)=-\frac{C_0}{2} \frac{(\Delta+1) u_1^{\Delta+1}}{(u-u_1)^{\Delta}(u u_1+1)^{\Delta+2}}$, and write
\be\label{IntDelV}
\Delta v =-\frac{C_0}{2} \int_{u_0}^{\infty} du_1 u_1^{\Delta+1} \int_{u_1}^{\infty} du\, \frac{(\Delta+1) }{(u-u_1)^{\Delta}(u u_1+1)^{\Delta+2}}\,.
\ee
Defining $w = \frac{u-u_1}{u+u_1}$, and after some manipulations, (\ref{IntDelV}) becomes
\bea
\Delta v &=&-\frac{(\Delta+1) C_0}{2^{\Delta}} \int_{u_0}^{\infty} du_1 u_1^2 \int_{0}^{1} dw \, \frac{(1-w)^{2 \Delta} w^{-\Delta}}{\left[ 1+u_1^2 +(u_1^2-1)w \right]^{\Delta+2}}\,, \nonumber\\
&=& -\frac{(\Delta+1) C_0}{2} \frac{\Gamma(1-\Delta) \Gamma(1+2\Delta)}{\Gamma(\Delta+2)} \int_{u_0}^{\infty} du_1 \frac{u_1^{2 \Delta}}{(1+u_1)^{2\Delta+1}}\,.
\eea
Next, we use the integral representation of the incomplete Beta function
\be
\int_{u_0}^{\infty} du_1 \frac{u_1^{2 \Delta}}{(1+u_1^2)^{2\Delta+1}} =- \frac{i (-1)^{-\Delta}}{2} B(-u_0^{-2};\Delta+1/2 ,-2\Delta)\,,
\ee
together with the following identities
\bea
B(x;a,b) &=& \frac{x^a}{a} {}_2 F_1(a,1-b, a+1;x)\,,\\
{}_2F_1(A,B,C;x)&=&(1-x)^{-A} {}_2 F_1 \left(A,C-B,C; \frac{x}{x-1} \right)\,,\\
 \sin(\pi \Delta)&=&\frac{\pi}{\Gamma(1-\Delta) \Gamma(\Delta)}\,,
\eea
to finally show that
\be
\Delta v = - h \ell_s^2 \frac{ \pi}{2\Delta+1} \frac{\Gamma(2\Delta+1)}{\Gamma(\Delta)^2} c_{\Delta}^2 \frac{{}_2 F_1\left( \Delta+1/2,1/2-\Delta, \Delta+3/2,\frac{1}{1+u_0^2}\right)}{(1+u_0^2)^{\Delta+1/2}}\,.
\ee
The above result was obtained for $h(t)=h \,\theta(t-t_0)$ and $r_0=1$. For the case of a instantaneous perturbation, i.e., $h^{\text{inst}}(t)= h \,\delta(t-t_0)$, we obtain\footnote{Note that $\delta(t-t_0)= - \partial_{t_0} \theta(t-t_0)$, and $\partial_{t_0} = u_0 \partial_{u_0}$. This implies that $\Delta v^{\text{inst}}=-u_0 \partial_{u_0} \Delta v (u_0)$. }
\be \label{eq-shiftinst}
\Delta v^{\text{inst}}=-u_0 \partial_{u_0} \Delta v (u_0) = - \pi h \ell_s^2 \frac{\Gamma(2\Delta+1)}{\Gamma(\Delta)^2} c_{\Delta}^2 \left( \frac{u_0}{1+u_0^2} \right)^{2 \Delta+1}\,.
\ee
We recall that in the above formulas $u_0=e^{t_0}$.

\subsection*{Case II: $\delta H(t) = h(t)\,  \dot{\phi}_L(-t) \dot{\phi}_R(t)$ }
In this case, we obtain
\be
G_h = 2 \sin (\pi \Delta) \int dt_1 h(t_1) \tilde{K}(t'+t_1-i \beta/2,r')\tilde{K}^r(t-t_1,r)+(t \leftrightarrow t')\,,
\ee
where
\bea
\tilde{K}(t-t',r) & \equiv & \langle \phi(t,r) \dot{\phi}(t',\infty)\rangle \,, \\
\tilde{K}^r(t-t',r) & =& |\tilde{K}(t-t',r)| \theta(t) \, \theta \left( \frac{\sqrt{r^2-r_0^2}}{r_0} \cosh(t-t')+\frac{r}{ r_0}  \right)\,,
\eea
denote a Wightman bulk-to-boundary propagator and a retarded bulk-to-boundary propagator, respectively. The difference with respect to the previous case is that now the propagators involve a source term $\dot{\phi}(t,r=\infty)$ at the boundary, instead than the standard $\phi(t,r=\infty)$. Other than that, the calculation follows the steps of the case I studied above. Once again, we will carry out the analysis for a general dimension $\Delta$ and at the end, we will set $\Delta=1$. 

To start with the calculation we need a propagator of the form
\be
\tilde{K}_{\Delta}= \langle \phi_{\mathcal{O}}(r,t) \, \dot{\mathcal{O}}(t') \rangle\,,
\ee
where $\phi_{\mathcal{O}}(r,t)$ is the bulk field dual to the operator $\mathcal{O}$. We can obtain this propagator as
\be
\tilde{K}_{\Delta}(t,r;t')= \langle \phi_{\mathcal{O}}(t,r) \, \dot{\mathcal{O}}(t') \rangle = \partial_{t'} \langle \phi_{\mathcal{O}}(t,r) \, \mathcal{O}(t') \rangle= \partial_{t'} K_{\Delta}(t,r;t')\,,
\ee
where $K_{\Delta}(t,r;t') \equiv \langle \phi_{\mathcal{O}}(t,r) \, \mathcal{O}(t') \rangle$. Using the above identity, we write
\be
G_h = 2 \sin (\pi \Delta) \int dt_1 h(t_1) \partial_{t_1} K_{\Delta}(t'+t_1-i \beta/2,r')\partial_{t_1} K^r_{\Delta}(t-t_1,r)+(t \leftrightarrow t')\,,
\ee
or, in terms of Kruskal coordinates, 
\be
G_h = \tilde{C}_0 \int \frac{du_1}{u_1} h(u_1) \frac{u u'}{\left( \frac{u}{u_1}-1\right)^{\Delta+1} \left( 1+u' u_1 \right)^{\Delta+1}} +(u \leftrightarrow u')\,.
\ee
Here we used
\bea
\tilde{K}_{\Delta} &=& u_1 \partial_{u_1} K_{\Delta} = c_{\Delta} \Delta \frac{u}{u_1}  \left[ \frac{u}{u_1}-1\right]^{-\Delta-1},\\
\tilde{K}_{\Delta}^{r} &=& u_1 \partial_{u_1} K_{\Delta}^r = - c_{\Delta} \Delta u' u_1  \left[ u' u_1+1\right]^{-\Delta-1},
\eea
where
\bea
K_{\Delta}(u',0;-t_1+i \pi)&=& c_{\Delta} \left[ u' u_1 +1 \right]^{-\Delta}\,,\\
K_{\Delta}^{r}(u,0;t_1) &=& c_{\Delta} \left[ \frac{u}{u_1}-1\right]^{-\Delta}
\eea
and $u_1=e^{t_1}$. Note also that $\partial_{t_1} = u_1 \partial_{u_1}$.

Using $h(u_1)= h\, \theta(u_1-u_0)$ and proceeding as for the previous case, we obtain
\bea
\Delta v &=&\ell_s^2 h \frac{\pi \Gamma(2\Delta)}{\Gamma(\Delta)^2} (-u_0^{-2})^{1/2-\Delta} u_0^{1-2\Delta} \Delta (1+\Delta) \Big[ \Delta B\left(-u_0^{-2};1/2+\Delta,-2\Delta-2 \right)+\nonumber\\
&& 2(1+\Delta)B\left(-u_0^{-2};3/2+\Delta,-2\Delta-2 \right)  + \Delta B\left(-u_0^{-2};5/2+\Delta,-2\Delta-2 \right) \Big]\,,
\eea
where $B(x;a,b)$ is the incomplete Beta function. Likewise, the result for an instantaneous perturbation reads
\be \label{eq-caseIIres}
\Delta v^{\text{inst}}= -u_0 \partial_{u_0} \Delta v =2 \pi h \ell_s^2 \frac{\Delta (\Delta+1) \Gamma( 2 \Delta)c_{\Delta}^2}{\Gamma(\Delta)^2} \frac{u_0^{2\Delta+1}}{(1+u_0^2)^{2\Delta+3}} \Big[\Delta-2(\Delta+1)u_0^2+\Delta u_0^4 \Big]\,.
\ee

Interestingly, the above result for case II can be obtained from the case I result as follows. First, we consider a slightly more general type I deformation: $\delta H_I = h \phi_L(t_L) \phi_R (t_R)$. Following Sec.~2.2 of \cite{Freivogel:2019whb}, we can see that the wormhole opening in this boosted frame reads:\footnote{In particular, this expression corresponds to the 2-dimensional version of Eq. (2.21) in \cite{Freivogel:2019whb}.}
\be
\Delta v_I(t_L,t_R) = \frac{\pi c_{\Delta}^2 \Gamma(2\Delta+1)}{2^{2+4\Delta}\Gamma(\Delta)^2} \exp\left(-\frac{\pi}{\beta}(t_L+t_R) \right)\left[ \frac{1}{2} \cosh\left( \frac{\pi}{\beta}(t_R-t_L)\right)\right]^{-2\Delta-1},
\ee
where we have considered that both fields $\phi_L$ and $\phi_L$ have dimension $\Delta$. Repeating the same calculation for slightly more general type II perturbations of the form $\delta H_{II} = h \dot{\phi}_L(t_L) \dot{\phi}_R (t_R) = h \partial_{t_L} \partial_{t_R}  \phi_L(t_L) \phi_R (t_R)$, it is easy to see that
\be
\Delta v_{II}(t_L,t_R) = \partial_{t_L} \partial_{t_R} \Delta v_I (t_L,t_R,\Delta)\,.
\ee
Similarly,  we can check that (\ref{eq-caseIIres}) can be obtained as $\partial_{t_L} \partial_{t_R} \Delta v_I (t_L,t_R,\Delta)\big|_{t_R=t_0, t_L=-t_0}$ and using that $u_0=e^{r_0 t_0}$.

\begin{figure}[t!]
\begin{center}
\begin{tabular}{cc}
\setlength{\unitlength}{1cm}
\hspace{-0.9cm}
\includegraphics[width=7.5cm]{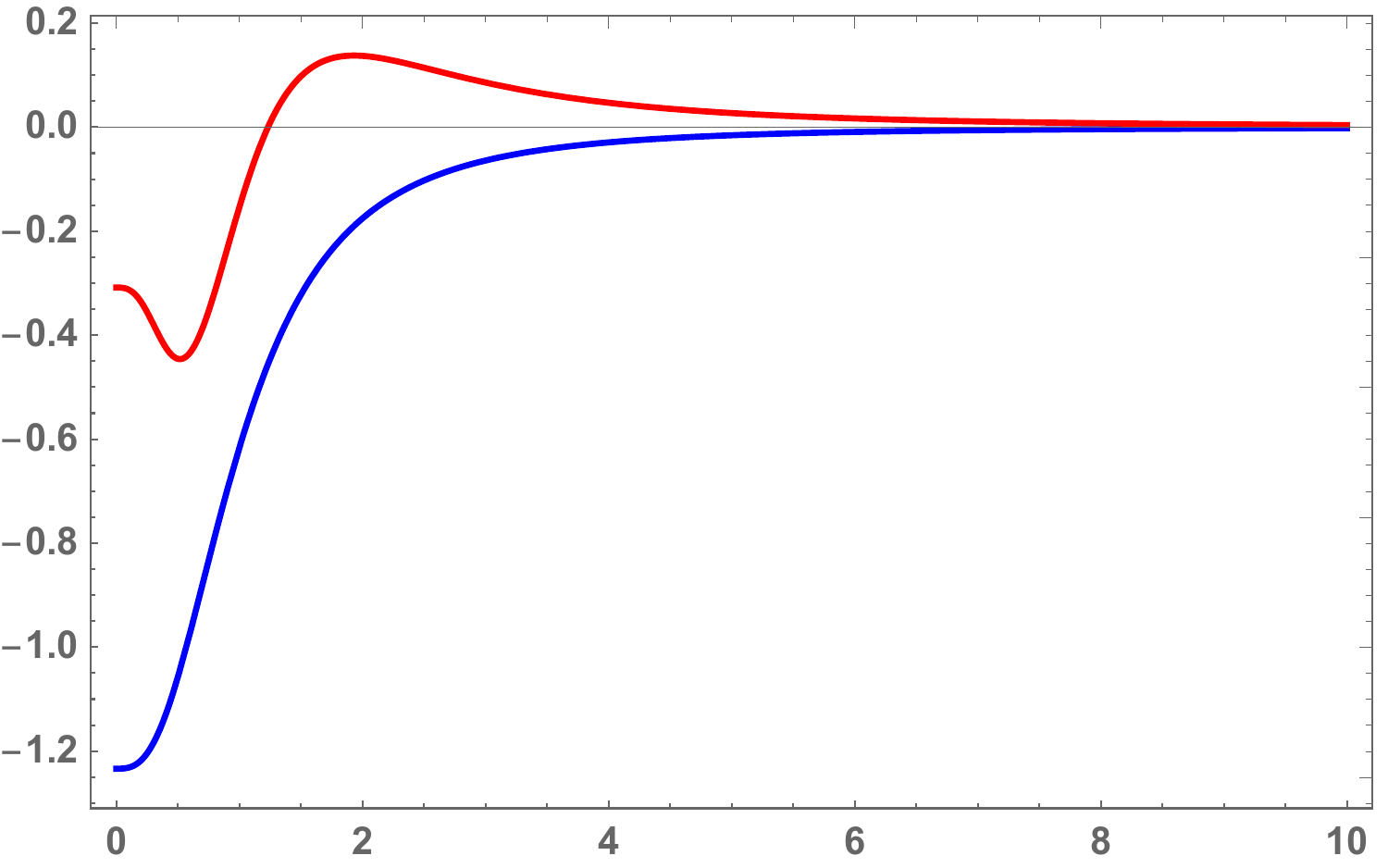}
\qquad\qquad &
\includegraphics[width=7.5cm]{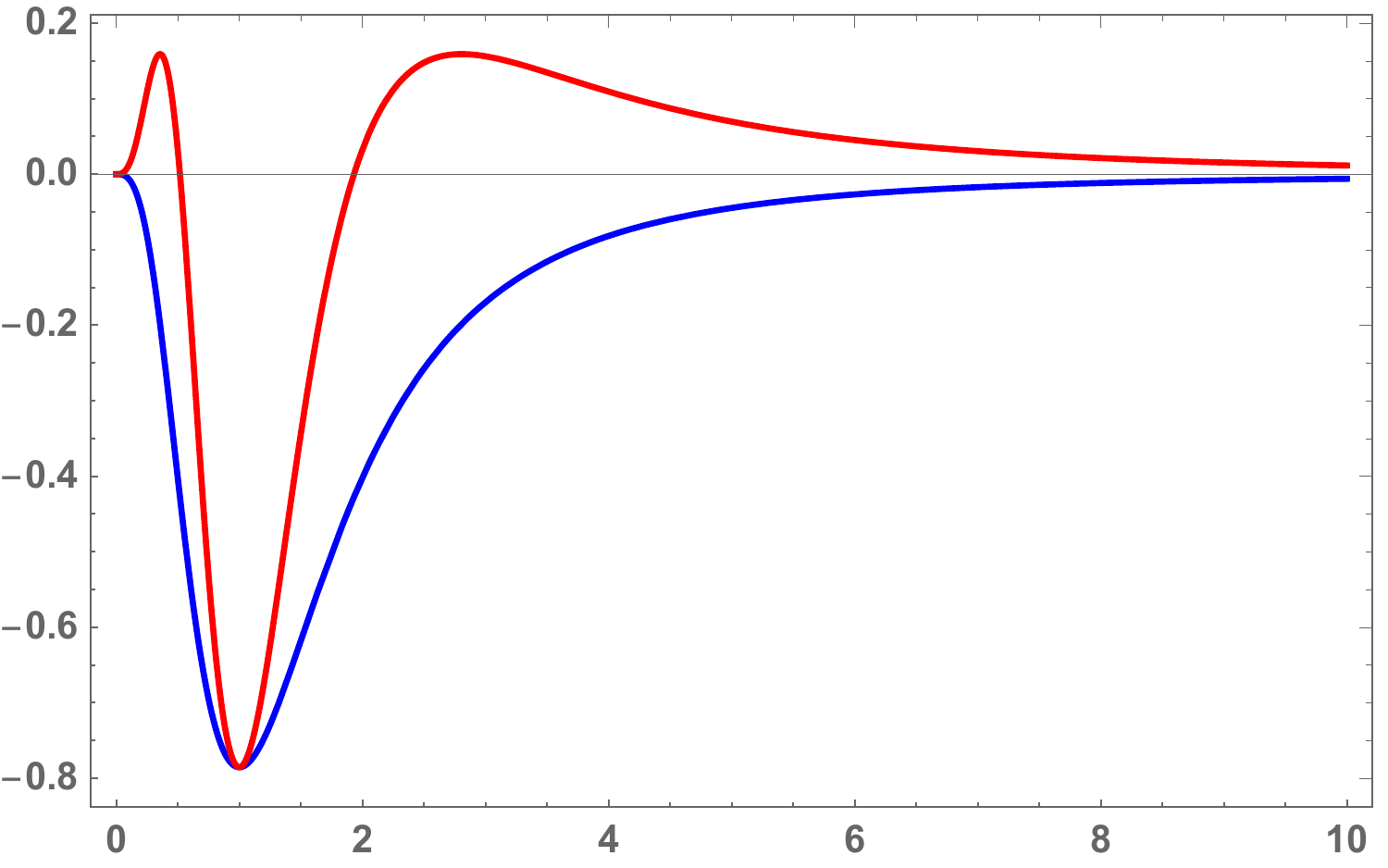}
\qquad
\put(-100,-10){$u_0$}
\put(-355,-10){$u_0$}
\put(-480,60){\rotatebox{90}{$\Delta v$}}
\put(-228,60){\rotatebox{90}{$\Delta v^\text{inst}$}}
\end{tabular}
\end{center}
\caption{$\Delta v$ versus $u_0=e^{t_0}$ for the cases I (blue curves) and II (red curves). Here, we set $h=\ell_s^2=1$. We fix $\Delta=1$ because the double trace deformation involves a massless $\phi$, which is dual to a dimension one operator. The left panel shows the result for $h(t)=h \theta(t-t_0)$, while the right panel shows the result for $h(t) = h \delta(t-t_0)$. }
    \label{fig:ANEC}
\end{figure}

In Fig.~\ref{fig:ANEC}, we set $\Delta=1$ and plot $\Delta v$ as a function of $u_0$ for type I and type II deformations. The left panel shows the result for a deformation that is turned on at $t_0$ and remains turned on forever. The right panel shows the result for an instantaneous perturbation. While in case I $\Delta v$ is negative for any $u_0$, in case II $\Delta v$ is only negative inside some interval $u_0 \in (u_1,u_2)$. In both cases, traversability for an instantaneous perturbation is optimal for $u_0=1$, which corresponds to $t_0=0$.

\subsection{Two-sided commutator} \label{sec-commutator}

In this section, we will diagnose traversability by sending a signal through the wormhole. The signal is produced by some generic boundary operator $W$, which is dual to a bulk field $\phi_W$. Under certain conditions, the $W$-quanta traverses the wormhole, producing a non-zero two-sided correlator of the form
\be
\langle [W_L(t_L), e^{-i \mathcal{V}} W_R(t_R) e^{i \mathcal{V}}] \rangle\,,
\ee
where, for the two cases we consider, the double trace deformation is given by
\bea \label{eq-deformationV}
\mathcal{V}_\text{I}= \frac{1}{K}\sum_{i=1}^{K} \int dt \,h(t) V^i_L(-t) V^i_R(t)\,,\\
\mathcal{V}_\text{II}= \frac{1}{K}\sum_{i=1}^{K} \int dt \,h(t) \dot{V}^i_L(-t) \dot{V}^i_R(t)\,,
\eea
and $V^i$ are boundary operators. We use $K$ different fields because the large-$K$ limit leads to simplifications, e.g., it suppresses particle creation in the worldsheet and enhances the effects of the coupling.\footnote{A concrete suggestion was given in \cite{Maldacena:2017axo} (see Appendix C.2. for details). In detail, the idea was to put $K$ strings at the north pole of the $S^3$ and $K$ oppositely oriented strings at the south pole of $S^3$. This is a BPS configuration, related to 1/2 BPS Wilson loops, so it is stable. In general, it is actually rather non-trivial to find microscopic realizations of large $K$ as these typically involve multiple objects with an emergent $U(K)$ gauge symmetry while the non-local deformation is not gauge invariant under this $U(K)$ gauge group.} The function $h(t)$ specifies the profile of the deformation. For example, for an instantaneous perturbation that is turned on at time $t=t_0$, we use $h(t)= h\, \delta(t-t_0)$. The Gao-Jafferis-Wall setup is obtained when we use a Heaviside step function, i.e., $h(t)= h\, \theta(t-t_0)$. 

Taking $W$ and $V^i$ as Hermitian operators, we can write
\be
\langle [W_L(t_L), e^{-i \mathcal{V}} W_R(t_R) e^{i \mathcal{V}}] \rangle = - 2\,i\, \text{Im}\, C\,,
\ee
where
\be
C = \langle  e^{-i \mathcal{V}} W_R(t_R) e^{i \mathcal{V}} W_L(t_L) \rangle\,.
\ee
In the large-$K$ and small-$\ell_s$ limit, we can write \cite{Maldacena:2017axo}
\be
C= e^{-i\langle \mathcal{V}\rangle} \tilde{C}, \,\,\,\, \tilde{C}= \langle W_R(t_R) e^{-i \mathcal{V}} W_L(t_L) \rangle\,.
\ee
In the following, we will consider the two cases separately.

\subsection*{Case I:}
Let us start by considering a deformation of case I. Using the eikonal approximation, we can write $\tilde{C}$ as
\be
\tilde{C}= \alpha \int dp^{u} p^{u} \psi_2^{*}(p^u)\psi_3(p^u)e^{i D}\,,
\ee
with
\be \label{eq-defD}
D  = \alpha \int dt\, h(t) \int dp^v p^{v}\psi_1^{*}(p^v)\psi_4(p^v)\,e^{i \delta(p^u p^v)}\,,
\ee
and $\alpha=\frac{4}{\pi r_0}$. The wave functions $\psi_i$ are given by
\bea
\psi_1(p^v)&=&\int du e^{2i p^v u} \langle \phi_{V}(u,v) V_L(-t) \rangle|_{v=0},\\
\psi_2(p^u)&=&\int dv e^{2i p^u v} \langle \phi_{W}(u,v) W_R(t_R) \rangle|_{u=0},\\
\psi_3(p^u)&=&\int dv e^{2i p^u v} \langle \phi_{W}(u,v) W_L(t_L) \rangle|_{u=0},\\
\psi_4(p^v)&=&\int du e^{2i p^v u} \langle \phi_{V}(u,v) V_R(t) \rangle|_{v=0},
\eea
where $\phi_V$ and $\phi_W$ are the bulk fields dual to the operators $V$ and $W$.
The phase shift describing the collision between the $W$-quanta and the $V$-quanta reads
\be
\delta(p^u p^v) = \ell_s^2 p^u p^v\,.
\ee
The bulk-to-boundary propagators for generic scalar operators are given by
\bea
 \langle \phi_{V}(u,v) V(t) \rangle &=& c_{\Delta_V} \left[ \frac{1+uv}{v e^{t}-u e^{-t}+(1-uv)} \right]^{\Delta_V},\\
  \langle \phi_{W}(u,v) W(t) \rangle &=& c_{\Delta_W} \left[ \frac{1+uv}{v e^t-u e^{-t}+(1-uv)} \right]^{\Delta_W},
\eea
where we assume that both $(u,v)$ and $t$ are points on the right-exterior region of the geometry. Fields on the left-exterior region can be obtained by replacing $t \rightarrow t + i \frac{\beta}{2}$ in the above formulas. The metric is $ds^2=4 du dv /(1+u v)^2$. In Schwarzschild coordinates, the horizon radius is $r_0$.

After performing the above integrals, we obtain
\bea
\psi_1&=&  2^{\Delta_V}  \frac{\pi r_0^{\Delta_V} e^{i \pi \Delta_V /2}}{ \Gamma(\Delta_V)} e^{-r_0 t \Delta_V} e^{-2i p^{v}e^{-r_0 t}} \theta(p^v) (p^v)^{\Delta_V-1}\,,\\
\psi_2&=&  2^{\Delta_W}  \frac{ \pi  r_0^{\Delta_W} e^{i \pi \Delta_W /2}}{ \Gamma(\Delta_W)} e^{-r_0 t_R \Delta_W} e^{-2i p^{u}e^{-r_0 t_R}} \theta(p^u) (p^u)^{\Delta_W-1}\,,\\
\psi_3&=&  2^{\Delta_W}  \frac{\pi  r_0^{\Delta_W} e^{-i \pi \Delta_W /2}}{\Gamma(\Delta_W)} e^{-r_0 t_L \Delta_W} e^{-2i p^{u}e^{-r_0 t_L}} \theta(p^u) (p^u)^{\Delta_W-1}\,,\\
\psi_4&=&  2^{\Delta_V}  \frac{\pi  r_0^{\Delta_V} e^{-i \pi \Delta_V /2}}{\Gamma(\Delta_V)} e^{r_0 t \Delta_V} e^{2i p^{v}e^{r_0 t}} \theta(p^v) (p^v)^{\Delta_V-1}\,.
\eea
Let us now compute $D$. Using the above wave functions, we obtain
\bea
D&=&\alpha \int dt\, h(t) \int dp^v p^{v}\psi_1^{*}(p^v)\psi_4(p^v)\,e^{i \delta(p^u p^v)}\,, \nonumber\\
&=& \alpha_V \int dt\, h(t) \int dp^{v} (p^v)^{2\Delta_V-1} e^{4 i p^{v} \cosh (r_0 t)} e^{i \ell_s^2 p^u p^v}\,, \nonumber\\
&=&  \alpha_V \Gamma(2\Delta_V) e^{i \pi \Delta_V} \int dt\, h(t) \left[ 4 \cosh(r_0 t)+ \ell_s^2 p^{u} \right]^{-2\Delta_V} \label{eq-D(p)},
\eea
where $\alpha_V = \pi^2 \alpha 2^{2\Delta_V}\frac{r_0^{2 \Delta_V}e^{-i \pi \Delta_V}}{ \Gamma(\Delta_V)^2} c_{\Delta_V}^2$ and we used that $\int dp p^{2\Delta-1} e^{i p F} = \Gamma(2\Delta)e^{i \pi \Delta} F^{-2\Delta}$, which is valid when Im($F$)$>0$.\footnote{This condition can be satisfied by giving a small imaginary piece to $t$.}

The correlator $\tilde{C}$ can then be written as
\be
\tilde{C} = \alpha_W e^{-r_0 \Delta_W (t_L+t_R)} \int dp^{u} (p^u)^{2\Delta_W-1}e^{2i p^{u} \left[ e^{-r_0 t_L}+e^{-r_0 t_R}  \right]} e^{i \alpha_V \Gamma(2\Delta_V) e^{i \pi \Delta_V} \int dt\, h(t) \left[ 4 \cosh(r_0 t)+ \ell_s^2 p^{u} \right]^{-2\Delta_V}}\!\!\!,
\ee
where $\alpha_W =\pi^2  \alpha 2^{2\Delta_W} \frac{r_0^{2 \Delta_W}e^{-i \pi \Delta_W}}{ \Gamma(\Delta_W)^2} c_{\Delta_W}^2$.
For simplicity, let us consider the case in which $h(t)=h r_0^{1-2\Delta} \delta(t-t_0)$. In this case, $\tilde{C}$ becomes
\be
\tilde{C} = \alpha_W e^{-r_0 \Delta_W (t_L+t_R)} \int dp^{u} (p^u)^{2\Delta_W-1}e^{2i p^{u} \left[ e^{-r_0 t_L}+e^{-r_0 t_R}  \right]} e^{i \alpha_V \Gamma(2\Delta_V) e^{i \pi \Delta_V}  h r_0^{1-2\Delta} \left[ 4 \cosh(r_0 t_0)+ \ell_s^2 p^{u} \right]^{-2\Delta_V}}\!\!\!.
\ee
The expectation value $\langle \mathcal{V} \rangle$ is obtained from $D$ with $\delta=0$, i.e.,
\be
\langle \mathcal{V} \rangle = D|_{\delta=0} = h r_0^{1-2\Delta} \alpha_V \Gamma(2\Delta_V) e^{i \pi \Delta_V} \left[ 4 \cosh(r_0 t_0) \right]^{-2\Delta_V}\,.
\ee
We then obtain
\be
\tilde{C} =\alpha_W e^{-r_0 \Delta_W (t_L+t_R)} \int dp^{u} (p^u)^{2\Delta_W-1}e^{2i p^{u} \left[ e^{-r_0 t_L}+e^{-r_0 t_R}  \right]} e^{i \langle \mathcal{V} \rangle   \left[ 1+ \frac{\ell_s^2 p^{u}}{4 \cosh (r_0 t_0)} \right]^{-2\Delta_V}},
\ee
and
\be \label{eq-Cnumerical}
C= e^{-i\langle \mathcal{V}\rangle} \tilde{C} = \alpha_W e^{-r_0 \Delta_W (t_L+t_R)} \int_{0}^{\infty} dp^{u} (p^u)^{2\Delta_W-1}e^{2i p^{u} \left[ e^{-r_0 t_L}+e^{-r_0 t_R}  \right]} e^{i \langle \mathcal{V} \rangle   \left[ \left( 1+ \frac{\ell_s^2 p^{u}}{4 \cosh (r_0 t_0)} \right)^{-2\Delta_V} -1\right] }\,.
\ee

We can study (\ref{eq-Cnumerical}) numerically to understand the properties of $C$, but it is also possible to obtain a closed analytic result in the probe approximation, which is obtained by considering a small-$p^u$ approximation, i.e., we consider $D= D_0 + D_1 p^u$, where $D_0 = \langle \mathcal{V} \rangle$: 
\bea
C_\text{probe}&=& \alpha_W e^{-r_0 \Delta_W (t_L+t_R)} \int_{0}^{\infty} dp^{u} (p^u)^{2\Delta_W-1}e^{2i p^{u} \left[ e^{-r_0 t_L}+e^{-r_0 t_R}  \right]} e^{2 i p^u \left[-\frac{\langle \mathcal{V} \rangle \Delta_V e^{i \pi \Delta_V} \ell_s^2 }{4 \cosh(r_0 t_0)} \right]  }\,,\nonumber\\
&=& \alpha_W e^{-r_0 \Delta_W (t_L+t_R)} \Gamma(2\Delta_W) e^{i \pi \Delta_W} \left[ 2 \left( e^{-r_0 t_L}+e^{-r_0 t_R} \right) + D_1  \right]^{-2\Delta_W}\,,
\eea
where
\be
D_1 = -\frac{ 2 h r_0^{1-2\Delta} \alpha_V \Delta_V \Gamma(2\Delta_V) e^{i \pi \Delta_V} \ell_s^2 }{\left(4 \cosh(r_0 t_0) \right)^{2\Delta_V+1}} c_{\Delta_V}^2\,.
\ee
Using the definitions of $\alpha$, $\alpha_W$ and $\alpha_W$, we can write
\bea \label{eq-probeI}
C_\text{probe}
&=& \pi 2^{2\Delta_W+2} c_{\Delta_W}^2 \frac{r_0^{2\Delta_W-1} \Gamma(2\Delta_W)}{ \Gamma(\Delta_W)^2}   \left[ 4 \cosh \left( r_0 \frac{t_L-t_R}{2} \right) + D_1 e^{r_0 \frac{t_R+t_L}{2}}  \right]^{-2\Delta_W}\!\!,
\eea
and
\be
D_1= -2 \pi h  \frac{ \Gamma(2\Delta_V+1)}{ \Gamma(\Delta_V)^2} \frac{\ell_s^2 c_{\Delta_V}^2}{(2\cosh r_0 t_0)^{2 \Delta_V+1}}\,,
\ee
where we use that $x\, \Gamma(x)=\Gamma(x+1)$.
The deformation changes the geodesic distance between the boundary points $t_L$ and $t_R$, and its effect is encoded in the term $D_1 e^{r_0 \frac{t_R+t_L}{2}}$. The term $D_1$ corresponds to a null shift in the $v$ direction, i.e., $D_1= 2 \Delta v$.

Using that $u_0=e^{r_0 t_0}$, the above result can be written as
\bea
\Delta v = \frac{D_1}{2}&=&  - \pi h  \frac{\Gamma(2\Delta_V+1)}{ \Gamma(\Delta_V)^2} \frac{\ell_s^2 c_{\Delta_V}^2}{(2\cosh r_0 t_0)^{2 \Delta_V+1}}\,, \\
&=& - \pi h  \frac{ \Gamma(2\Delta_V+1)}{ \Gamma(\Delta_V)^2}  \ell_s^2 c_{\Delta_V}^2 \left( \frac{u_0}{u_0^2+1} \right)^{2 \Delta_V+1}\,,
\eea
which perfectly agrees with the result (\ref{eq-shiftinst}) obtained via point-splitting for case I .
\subsection*{Case II}

Let us now compute the two-sided correlator for the second type of perturbation, which involves time derivatives of the boundary operators. We first compute $D$ as
\be
D= \alpha \int dt\, h(t) \int dp^v p^{v}\tilde{\psi}_1^{*}(p^v)\tilde{\psi}_4(p^v)\,e^{i \delta }\,,\\
\ee
where $\tilde{\psi}$ is the Fourier transform of $\tilde{K}$. These wave functions are obtained as
\bea
\tilde{\psi}_1(p^v) &=& \int du e^{2 i p^v} \tilde{K}(u,0;t)=\int du e^{2 i p^v} \partial_{t_1}K(u,0;t)=\partial_{t} \psi_1(p^v) \\
&=& i 2^{\Delta_V} c_{\Delta_V}^2 \frac{\pi r_0^{\Delta_V+1} e^{i \pi \Delta_V /2}}{ \Gamma(\Delta_V)} e^{-r_0 t \Delta_V} e^{-2i p^{v}e^{-r_0 t}} \theta(p^v) (p^v)^{\Delta_V-1} (2 i p^v e^{-r_0 t}-\Delta_V)\,,\nonumber \\
\tilde{\psi}_4(p^v)&=&i 2^{\Delta_V} c_{\Delta_V}^2 \frac{\pi r_0^{\Delta_V+1} e^{-i \pi \Delta_V /2}}{ \Gamma(\Delta_V)} e^{r_0 t \Delta_V} e^{2i p^{v}e^{r_0 t}} \theta(p^v) (p^v)^{\Delta_V-1} (2i p^v e^{r_0 t}+\Delta_V ) \,.
\eea
For simplicity, we set $r_0=1$ and consider an instantaneous perturbation $h(t)=h \delta(t-t_0)$. Using the above wave functions and $\delta= \ell_s^2 p^u p^v$, we can write
\bea
D &=&-\frac{2}{\pi} h  c_{\Delta_V}^2 2^{2\Delta_V}\frac{\pi^2}{\Gamma(\Delta_V)^2} e^{i \pi \Delta_V}\int dp^v (p^v)^{2\Delta_V-1} e^{4 i p^v \cosh t_0} e^{i \ell_s^2 p^u p^v} (2i p^v e^{r_0 t}+\Delta_V ) (-2i p^v e^{r_0 t}-\Delta_V ) \nonumber \\
&=&- h 4^{\Delta_V+1} \pi \frac{\Delta_V \Gamma(2\Delta_V)}{\Gamma(\Delta_V)^2} \frac{\left[8+(8+\ell_s^4 p^{u 2})\Delta_V-8\Delta_V \cosh(2t_0) \right]}{\left( 4 \cosh t_0+\ell_s^2 p^u\right)^{2\Delta_V+2}}\,.
\eea
Having obtained $D$, we can compute the expectation value $\langle \mathcal{V} \rangle$ as
\be
\langle \mathcal{V} \rangle = D|_{\delta=0} = \frac{2^{1-2\Delta_V} h \pi \Delta_V}{(\cosh t_0)^{2\Delta_V}} \frac{\Gamma(2\Delta_V)}{\Gamma(\Delta_V)^2} \left[ \Delta_V \cosh(2t_0)-\Delta_V-1 \right] \text{sech}(t_0)^2\,.
\ee
We can then compute the two-sided commutator as
\be \label{eq-Cnumerical2}
C= e^{-i \langle \mathcal{V} \rangle} \tilde{C} = \alpha_W e^{-\Delta_W (t_L+t_R)} \int_{0}^{\infty} dp p^{2\Delta_W-1} e^{2 i p \left(e^{-t_L}+e^{-t_R} \right)} e^{\langle \mathcal{V} \rangle \left[ \frac{\left(1+\frac{\Delta_V \ell_s^4 p^{u 2}}{8 [1+\Delta_V-\Delta_V \cosh (2t_0)]} \right)}{(1+\frac{\ell_s^2 p^u}{4 \cosh t_0})^{2\Delta_V+2}}-1\right]}\,.
\ee
The above expression can be used to study the behavior of $C$ numerically. Before doing that, let us first study $C$ in the probe limit, where we perform a small-$p^u$ approximation, $D= D_0 + D_1 p^u$, with $D_0 = \langle \mathcal{V} \rangle$. We first write
\be \label{eq-Cprobe22}
C_\text{probe}= \alpha_W e^{-\Delta_W (t_L+t_R)} \int_{0}^{\infty} dp p^{2\Delta_W-1} e^{2 i p \left(e^{-t_L}+e^{-t_R} \right)} e^{D_1 p^u}\,,
\ee
where
\bea \label{eq-D1case2}
D_1 &=& -4 h \ell_s^2 c_{\Delta_V}^2 \frac{\pi 2^{2\Delta_V}}{\Gamma(\Delta_V)^2} \frac{(\Delta_V+1)\Gamma(2\Delta_V+1)}{(4\cosh t_0)^{2\Delta_V+3}} \big(8-16 \Delta \sinh^2t_0 \big)\,, \\
&=&  4 \pi \ell_s^2 h \frac{\Gamma(2\Delta_V)\Delta_V (\Delta_V+1)}{\Gamma(\Delta_V)^2} c_{\Delta_V}^2 \frac{u_0^{2\Delta_V+1}}{\big( 1+u_0^2\big)^{2\Delta_V+3}} \left[2u_0^2(\Delta_V+1)-\Delta_V u_0^4-\Delta_V \right]\,,\nonumber
\eea
and $u_0=e^{t_0}$. Performing the integral in $p^u$ in (\ref{eq-Cprobe22}), we find
\be \label{eq-probeII}
C_\text{probe}= \frac{4 \pi \Gamma(2\Delta_W)}{\Gamma(\Delta_W)^2}c_{\Delta_W}^2 \left[ 2 \cosh\left( \frac{t_L-t_R}{2}\right)+\frac{D_1}{2}e^{\frac{t_R+t_L}{2}}\right]^{-2\Delta_W},
\ee
which has the same form as the two-sided commutator for case I, with the only difference that now $D_1$ is given by (\ref{eq-D1case2}). From $D_1$, we can extract the wormhole opening as
\be
\Delta v =\frac{D_1}{2}=  2 \pi \ell_s^2 h \frac{\Gamma(2\Delta_V)\Delta_V (\Delta_V+1)}{\Gamma(\Delta_V)^2} c_{\Delta_V}^2 \frac{u_0^{2\Delta_V+1}}{\big( 1+u_0^2\big)^{2\Delta_V+3}} \left[2u_0^2(\Delta_V+1)-\Delta_V u_0^4-\Delta_V \right]\,,
\ee
which perfectly matches the result obtained via point splitting (\ref{eq-caseIIres}).

Note that $\text{Im} (C_\text{probe}) = 0$ for integer values of $\Delta_W$, both for (\ref{eq-probeI}) and (\ref{eq-probeII}). This implies that, in the probe limit, we cannot send a message through the wormhole using $\phi$ or $\dot{\phi}$ as the signal. This is an artifact of the probe limit. We will see in the next section that is possible to obtain $\text{Im} (C) \neq 0$ if we include backreaction effects.

\subsubsection{Going beyond the probe approximation} \label{sec-beyondprobe}
In this section, we consider the first corrections beyond the probe approximation. We first note that the commutator given in (\ref{eq-Cnumerical}) or (\ref{eq-Cnumerical2}) has the form
\be \label{eq-Cw}
C = \alpha_W \int_{0}^{\infty} dP P^{2\Delta_W-1} e^{4 i P} e^{i(D(P)-D_0) }\,,
\ee
where we have set $r_0=1$, $c_{\Delta_V}=c_{\Delta_W}=1$ and $t_R=t_L=t$, and defined $P\equiv p^u e^{-t}$.

\begin{figure}[t!]
\centering
\includegraphics[width=7.5cm]{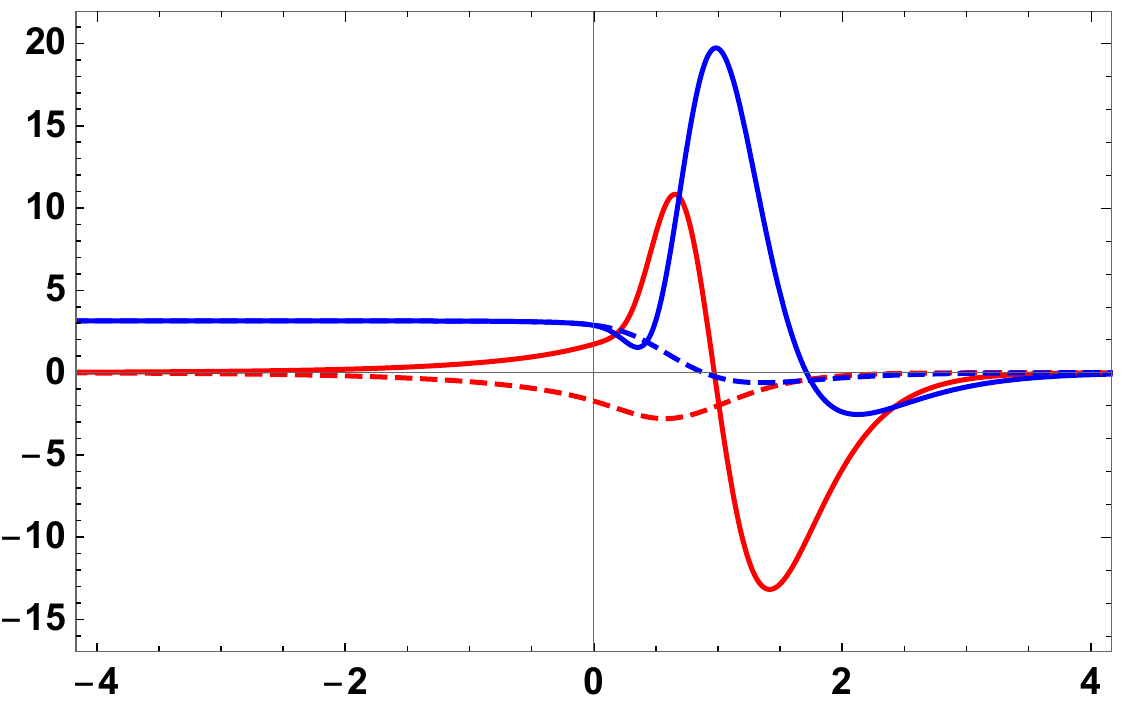}
\put(-190,80){ Re($C$)}
\put(-190,53){ Im($C$)}
\put(-120,-10){$t+\log(\ell_s^2)$}

 \caption{ Re($C$) (in blue) and Im($C$) (in red) versus $t+\log(\ell_s^2)$. The continuous (dashed) curves represent case I (II). Here we set $h=0.6$, $\Delta_V=\Delta_W=1$ and $t_0=0$. \label{fig-Cnonprobe}}
\end{figure}

We now expand $D(P)$ as
\be \label{eq-D2}
D(P)= D_0+ D_1 P + D_2 P^2+\cdots\,.
\ee
The piece $D_0+ D_1 P$ gives the probe limit result, while $D_2 P^2$ takes into account the first effects of backreaction. Substituting (\ref{eq-D2}) into (\ref{eq-Cw}) and performing the integral over $P$, we obtain
\bea \label{eq-CD2}
C &=& \alpha_W \frac{(-i D_2)^{\Delta_W}}{2}\Big[ \Gamma(\Delta_W)\, {}_1F_1\Big(\Delta_W,1/2,-i\frac{(4+D_1)^2}{4D_2}\Big)\nonumber\\
&+&\frac{(4+D_1)D_2}{(-i D_2)^{3/2}}\,\Gamma(\Delta_W+1/2)\, {}_1F_1\Big(\Delta_W+1/2,3/2,-i\frac{(4+D_1)^2}{4D_2}\Big) \Big] \,.
\eea
After specifying $D_1$ and $D_2$ for deformations of the cases I and II, the above equation can be used to study the time-dependence of the commutator.\footnote{Note that $D_1$ and $D_2$ are functions of $t$.}
Fig.~\ref{fig-Cnonprobe} shows the real and imaginary part of the commutator $C$ obtained with (\ref{eq-CD2}).

Now we study the full two-sided commutator by numerically evaluating the integrals (\ref{eq-Cnumerical}) for case I and (\ref{eq-Cnumerical2}) for case II. In both cases, we set $r_0=1$, $c_{\Delta_V}=c_{\Delta_W}=1$ and $t_R=t_L=t$. By  introducing $P=p^u e^{-t}$, the expressions for the two-sided correlators become
\bea
C_\text{I} &=& \alpha_W \int_{0}^{\infty} dP P^{2\Delta_W-1} e^{4 i P} e^{i D_0 \left[(1+\frac{\ell_s^2 P e^{t}}{4 \cosh t_0})^{-2\Delta_V} -1\right] }\,, \nonumber\\
D_0 &=& 4 \pi h \frac{ \Gamma(2\Delta_V)}{\Gamma(\Delta_V)^2} \left[2 \cosh( t_0) \right]^{-2\Delta_V}\,,
\eea
and
\bea
C_\text{II} &=& \alpha_W \int_{0}^{\infty} dP P^{2\Delta_W-1} e^{4 i P} e^{i D_0 \left[\left(1+\frac{\ell_s^2 P e^{t}}{4 \cosh t_0} \right)^{-2\Delta_V}\left( 1+\frac{\Delta_V \ell_s^4 P^2 e^{2t}}{8 [1+\Delta_V-\Delta_V \cosh (2t_0)]} \right) -1\right] } \,,\nonumber \\
D_0 &=&\frac{2 \pi h \Delta_V}{(2 \cosh t_0)^{2\Delta_V}} \frac{\Gamma(2\Delta_V)}{\Gamma(\Delta_V)^2} \left[ \Delta_V \cosh(2t_0)-\Delta_V-1 \right] \text{sech}(t_0)^2\,,
\eea
respectively.

\begin{figure}[t!]
    \centering
    \begin{minipage}{0.5\textwidth}
        \centering
        \includegraphics[width=0.9\textwidth]{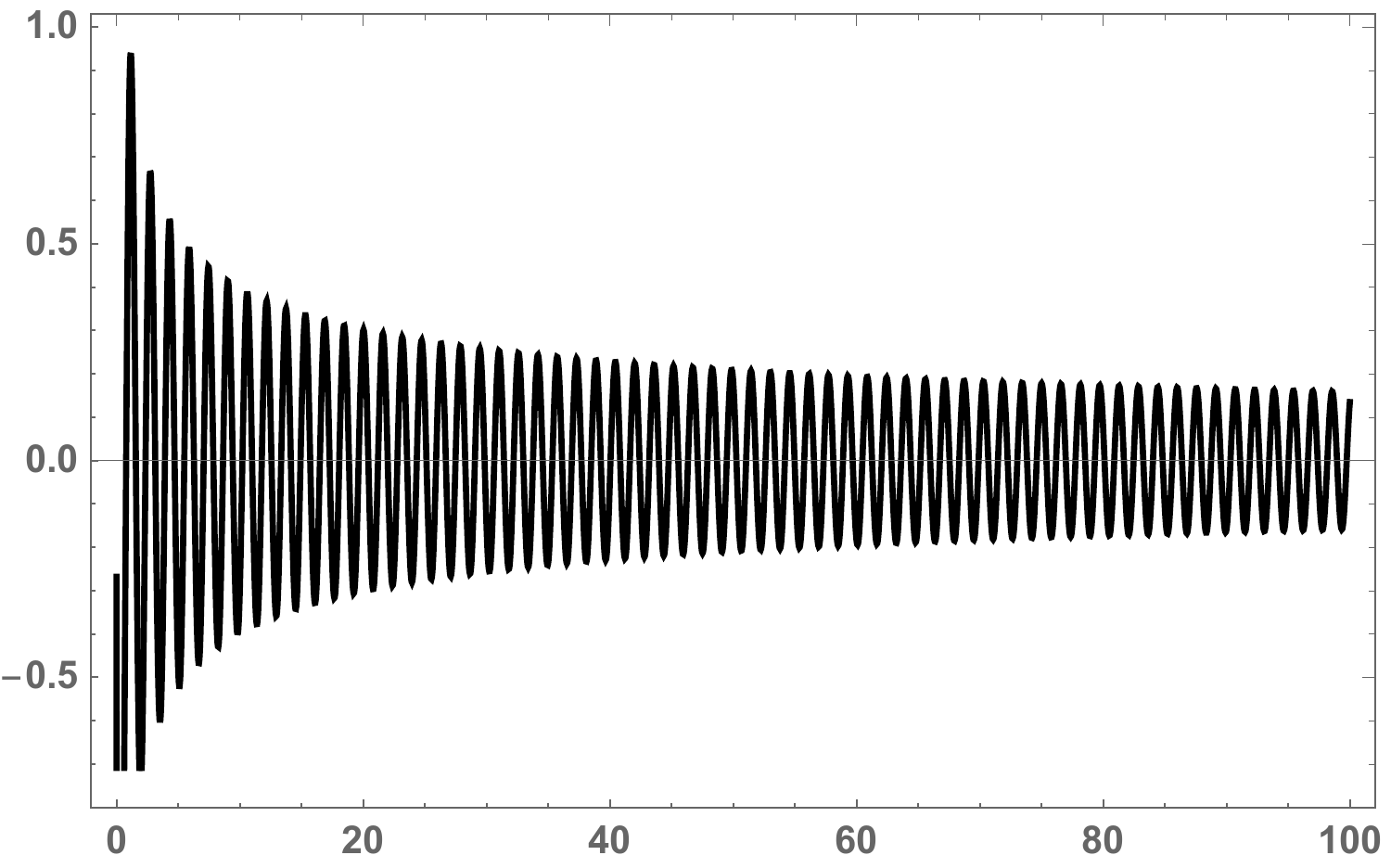} 
        \caption*{$\Delta_W =.3$}
    \end{minipage}\hfill
    \begin{minipage}{0.5\textwidth}
        \centering
        \includegraphics[width=0.9\textwidth]{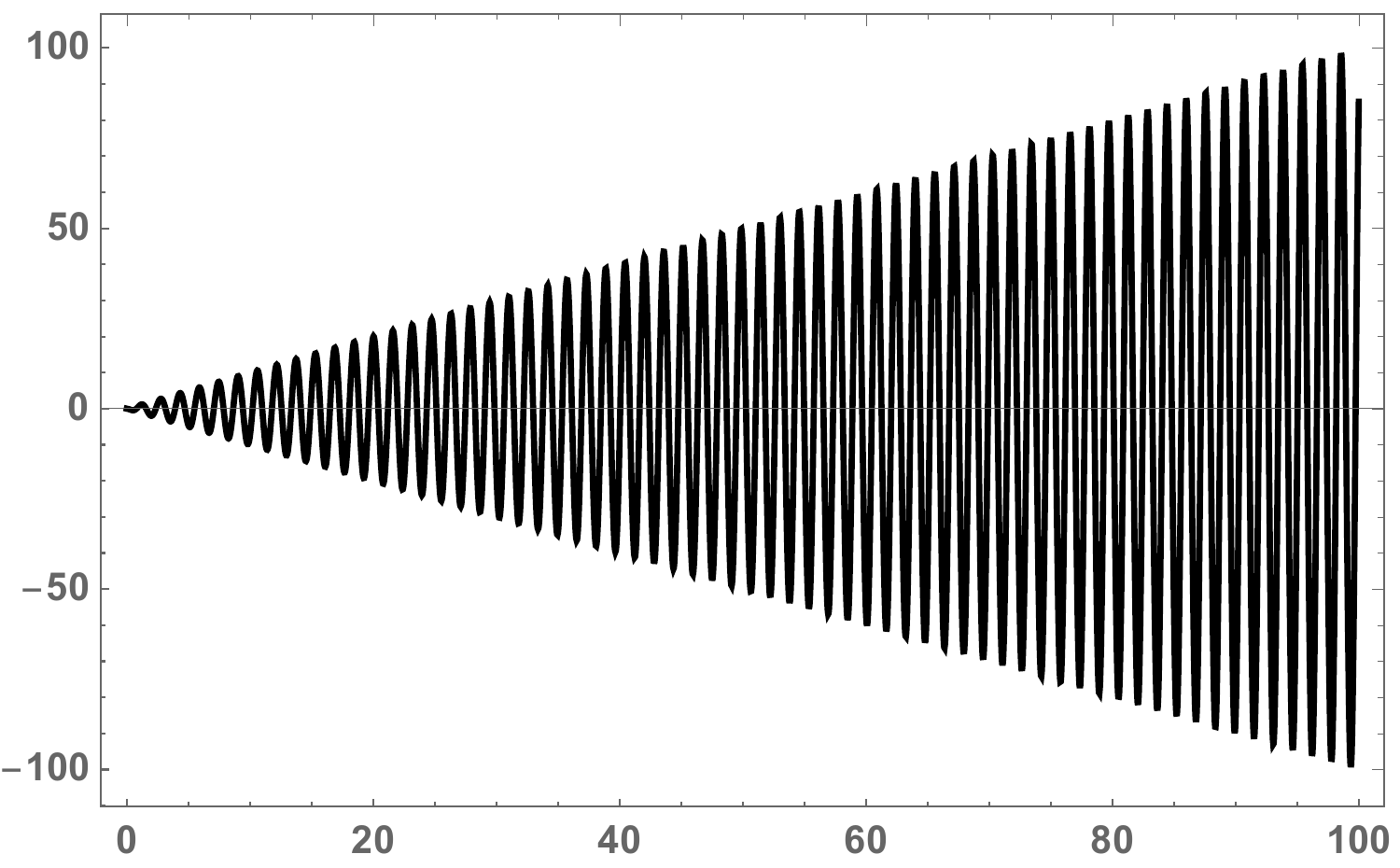} 
        \caption*{$\Delta_W =1$}
    \end{minipage}
    \caption{Integrand of $C = \alpha_W \int_{0}^{\infty} dP P^{2\Delta_W-1} e^{4 i P} e^{i(D(P)-D_0) }$  as a function of $P$ for $\Delta_W < 1/2$ (left panel) and $\Delta_W > 1/2$ (right panel). Here we consider a deformation of case I and we set $\Delta_V=1$, $\ell_s=h=c_V=c_W=1$, $t_0=0$ and $t=5$. \label{fig-integrand}}
\end{figure}

\begin{figure}[t!]
\centering
\includegraphics[width=3in]{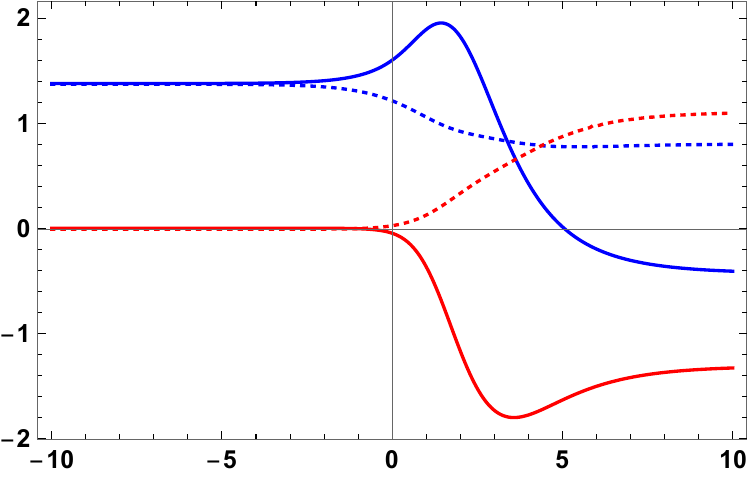}
\put(-190,118){ Re($C$)}
\put(-190,76){ Im($C$)}
\put(-120,-10){ $t+\log(\ell_s^2)$}

 \caption{ Re($C$) (in blue) and Im($C$) (in red) versus $t+\log(\ell_s^2)$. The continuous (dashed) curves represent case I (II). Here we set $h=0.6$, $\Delta_V=1$, $\Delta_W=0.3$ and $t_0=0$. \label{fig-Cnumerical1}}

\end{figure}

\subsubsection{Irrelevant deformations and imprint on the UV}

The double trace deformation (\ref{eq-deformationV}) is relevant for $\Delta_V \leq 1/2$. Since we are taking $V=\phi$ or $V=\dot{\phi}$, and $\phi$ is a massless field, this condition is not satisfied in our setup, and therefore the deformation will have a large imprint on the near boundary geometry. This could explain why in certain cases, the result for the commutator $C$ explodes as one considers high-$P$ contributions of the signal. In fact, insights coming from 2d dilaton gravities and $T\bar{T}$-deformed theories suggest that one of the leading effects of the irrelevant deformation may be the introduction of a natural UV cutoff scale that could cap off the problematic high-$P$ modes. 

To analyze this problem in more detail, we will simply introduce a UV cutoff in our calculation, and study its imprint on the observables to determine under what conditions our results are reliable. We introduce a UV cutoff by defining a maximal value $r_c$ for the AdS radial coordinate $r$.\footnote{This cutoff can be naturally identified as the position where the flavor branes end.} See Fig.~\ref{fig-PenroseCutoff}.
Once we introduce this cutoff, we need re-derive our bulk-to-boundary propagators $K$ since they were obtained for a spacetime with boundary at $r \rightarrow \infty$. We start by writing the expression for bulk-to-bulk propagators of a scalar field
\bea
G_{\Delta}(t,r;t',r')&=& c_{\Delta} \xi^{\Delta} {}_2F_1\left(\frac{\Delta}{2},\frac{\Delta+1}{2},\Delta+\frac{1}{2};\xi^{2}\right)\,,\\
\xi &=&\left[ - \sqrt{\frac{r^2}{r_0^2}-1}\sqrt{\frac{r'^2}{r_0^2}-1} \cosh \frac{ r_0 (t-t')}{\ell^2}+\frac{r r'}{r_0^2} \right]^{-1}\,,
\eea
which is not affected by the presence of the cutoff. Now, we compute the bulk-to-boundary propagator as
\bea \label{eq-Kcutoff}
K_c(t,r;t',r_c)&=& \lim_{r' \rightarrow r_c} r_c^{\Delta} G(t,r;t',r') \\
&=& c_{\Delta} \left[ - \sqrt{\frac{r^2}{r_0^2}-1}\, \sqrt{r_0^{-2}-r_c^{-2}} \cosh \frac{r_0(t-t')}{\ell^2}+\frac{r}{r_0^2} \right]^{-\Delta} \!\!\!\!\!\!\!{}_2F_1\left(\frac{\Delta}{2},\frac{\Delta+1}{2},\Delta+\frac{1}{2};\xi_c^{2}\right), \nonumber
\eea
where
\be
\xi_c = \frac{r_0}{r_c} \left[ - \sqrt{\frac{r^2}{r_0^2}-1}\sqrt{1-\frac{r_0^2}{r_c^2}} \cosh \frac{r_0(t-t')}{\ell^2}+\frac{r}{r_0} \right]^{-1}\,.
\ee
The strategy now is to compute the wormhole opening $\Delta v$ in terms of the propagator (\ref{eq-Kcutoff}) and study how the cutoff affects this quantity.
\begin{figure}[t!]
\centering

\begin{tikzpicture}[scale=1.5]
\draw [thick, dotted]  (0,0) -- (0,3);
\draw [thick,blue] (0,0) .. controls (.5,1) and (.5,2) .. (0,3);
\draw [thick,dotted]  (3,0) -- (3,3);
\draw [thick,blue] (3,0) .. controls (2.5,1) and (2.5,2) .. (3,3);
\draw [thick,dashed]  (0,0) -- (3,3);
\draw [thick,dashed]  (0,3) -- (3,0);
\draw [thick,decorate,decoration={zigzag,segment length=1.5mm, amplitude=0.3mm}] (0,3) -- (3,3);
\draw [thick,decorate,decoration={zigzag,segment length=1.5mm,amplitude=.3mm}]  (0,0) -- (3,0);

\draw[thick,<->] (1,2.2) -- (1.5,1.7) -- (2,2.2);

\end{tikzpicture}
\vspace{0.1cm}
\put(-140,50){\rotatebox{90}{\small $r = \infty$}}
\put(-124,50){\rotatebox{90}{\small $r = r_c$}}
\put(-2,80){\rotatebox{-90}{\small $r = \infty$}}
\put(-18,80){\rotatebox{-90}{\small $r = r_c$}}
\put(-75,-8){\small $r = 0$}
\put(-75,133){\small $r = 0$}
\put(-96,95){\small $v$}
\put(-46,95){\small $u$}

\caption{ \small Two-sided AdS black hole with a cutoff at $r=r_c$. The region near the asymptotic boundary at $r=\infty$ is removed, and the geometry ends at the cutoff surface (shown in blue) defined by $r=r_c$.}
\label{fig-PenroseCutoff}
\end{figure}

In Kruskal coordinates, $K_c$ can be written as
\be \label{eq-Kc}
K_c(u,v;t_1) = c_{\Delta} \left[ \frac{1+u v}{\sqrt{1-\eta^2}(v e^{r_0 t_1/\ell^2}-u e^{-r_0 t_1/\ell^2})+1-u v} \right]^{\Delta}{}_2F_1\left( \frac{\Delta}{2},\frac{\Delta+1}{2},\Delta+\frac{1}{2};\xi_c^{2} \right),
\ee
with
\be
\xi_c = \eta \left[ \frac{1+u v}{\sqrt{1-\eta^2}(v e^{r_0 t_1/\ell^2}-u e^{-r_0 t_1/\ell^2})+1-u v} \right],
\ee
and $\eta =\frac{r_0}{r_c}$.

\begin{figure}[t!]
\centering

\begin{tikzpicture}[scale=1.5]
\draw [thick, dotted]  (0,0) -- (0,3);
\draw [thick,blue] (0,0) .. controls (1.4,1.4) and (1.4,1.6) .. (0,3);
\draw [thick,dotted]  (3,0) -- (3,3);
\draw [thick,blue] (3,0) .. controls (1.6,1.4) and (1.6,1.6) .. (3,3);
\draw [thick,dashed]  (0,0) -- (3,3);
\draw [thick,dashed]  (0,3) -- (3,0);
\draw [thick,decorate,decoration={zigzag,segment length=1.5mm, amplitude=0.3mm}] (0,3) -- (3,3);
\draw [thick,decorate,decoration={zigzag,segment length=1.5mm,amplitude=.3mm}]  (0,0) -- (3,0);

\draw[thick,<->] (1,2.2) -- (1.5,1.7) -- (2,2.2);
\end{tikzpicture}
\vspace{0.1cm}
\put(-140,50){\rotatebox{90}{\small $r = \infty$}}
\put(-100,52){\rotatebox{90}{\small $r = r_c$}}
\put(-2,80){\rotatebox{-90}{\small $r = \infty$}}
\put(-42,78){\rotatebox{-90}{\small $r = r_c$}}
\put(-75,-8){\small $r = 0$}
\put(-75,133){\small $r = 0$}
\put(-96,95){\small $v$}
\put(-46,95){\small $u$}

\caption{ \small Two-sided AdS black hole with a cutoff at $r=r_c$ very close to the horizon. The region near the asymptotic boundary at $r=\infty$ is removed, and the geometry ends at the cutoff surface (shown in blue) defined by $r=r_c$.}
\label{fig-PenroseCutoff2}
\end{figure}
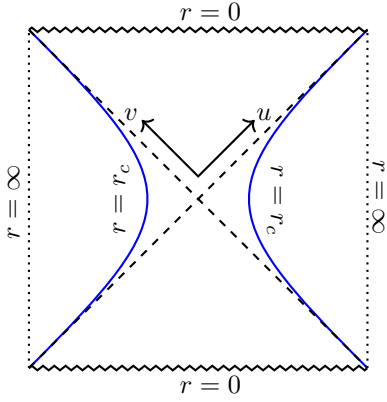

\begin{figure}[t!]
\centering
\includegraphics[width=4in]{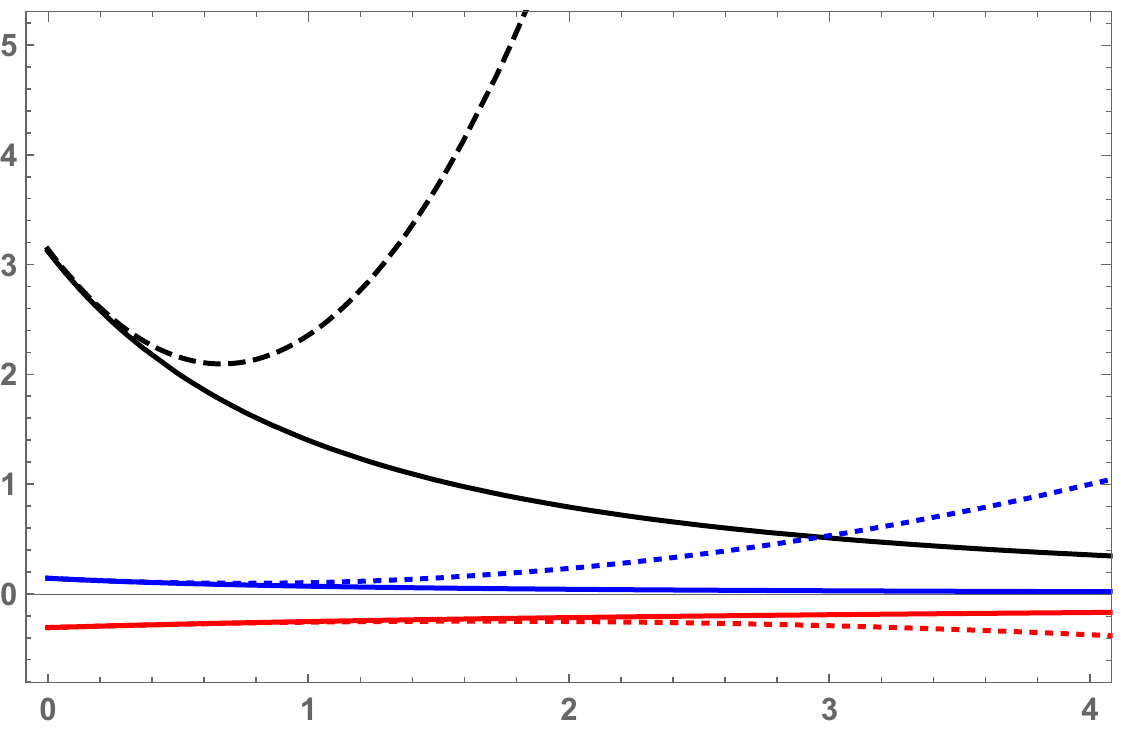}
\put(0,0){$P$}
\put(-320,100){$D(P)$}
\put(-270,120){\small $r_c=\infty$}
\put(-270,41){\small $r_c=100 r_0$}
\put(-270,17){\small $r_c=1.01 r_0$}

 \caption{ $D(P)$ (continuoues lines) as well as $D_0+D_1 P+D_2 P^2$ (dotted lines) for different values of the cutoff $r_c$. \label{fig-DPnumerical}}

\end{figure}

Let us first study the effect of the UV cutoff on $D$, defined in (\ref{eq-defD}). We start by writing $D$ in position space
\bea
D(p^u)&=& \alpha \int dt_1\, h(t_1) \int dp^v p^{v}\tilde{\psi}_1^{*}(p^v)\tilde{\psi}_4(p^v)\,e^{i \ell_s^2 p^v p^u }\,,\\
&=&\alpha \frac{\pi i}{2} \int \frac{du_1}{u_1}\, h(u_1) \int du K_1^*(u,0;t) \partial_u K_4\left(u-\frac{\ell_s^2}{2}p^u,0;t\right)\,, \label{eq-DwithK}
\eea
where
\bea
K_1\left(u,0;u_1\right) &=& c_{\Delta} \left[ \frac{1}{1 - (1-\eta)\frac{u}{u_1}} \right]^{\Delta}{}_2F_1\left( \frac{\Delta}{2},\frac{\Delta+1}{2},\Delta+\frac{1}{2};\left[ \frac{\eta}{1 - (1-\eta)\frac{u}{u_1}} \right]^2 \right),\nonumber\\
K_4(u-P,0;u_1)&=& c_{\Delta} \left[ \frac{1}{1 - (1-\eta)\frac{u-P}{u_1}} \right]^{\Delta}{}_2F_1\left( \frac{\Delta}{2},\frac{\Delta+1}{2},\Delta+\frac{1}{2};\left[ \frac{\eta}{1 - (1-\eta)\frac{u-P}{u_1}} \right]^2 \right),\nonumber
\eea
with $P=\frac{\ell_s^2}{2}p^u$, and $u_1 =\sqrt{\frac{r_c-r_0}{r_c+r_0}}e^{r_0 t_1/\ell^2}$. We introduce the variable $U=(1-\eta) u$, and define
\bea
k_1\left(U;u_1\right) &\equiv& K_1\left(\frac{U}{1-\eta},0;u_1 \right) = c_{\Delta} \left[ \frac{1}{1 - \frac{U}{u_1}} \right]^{\Delta}{}_2F_1\left( \frac{\Delta}{2},\frac{\Delta+1}{2},\Delta+\frac{1}{2};\left[ \frac{\eta}{1 - \frac{U}{u_1}} \right]^2 \right) , \nonumber\\
k_4(U;u_1)&\equiv& K_4\left(\frac{U}{1-\eta},0;u_1 \right)=c_{\Delta} \left[ \frac{1}{1 - \frac{U}{u_1}} \right]^{\Delta}{}_2F_1\left( \frac{\Delta}{2},\frac{\Delta+1}{2},\Delta+\frac{1}{2};\left[ \frac{\eta}{1 - \frac{U}{u_1}} \right]^2 \right)  ,\nonumber
\eea
in which terms $D(P)$ can be computed as
\be \label{eq-Dseries}
D(P)= \alpha \frac{\pi i}{2} \int \frac{du_1}{u_1}\, h(u_1) \int dU k_1^*(U,0;u_1) \partial_U k_4\left(U-(1-\eta)P,0;u_1\right)\,.
\ee
Note that $D$ depends on $P$ through the combination $(1-\eta)P$. Let us now consider a cutoff surface that is located very close to the horizon, such that $\epsilon=(1-\eta)=\frac{r_c-r_0}{r_c}$ is a small parameter. The corresponding Penrose diagram is illustrated in Fig.~\ref{fig-PenroseCutoff2}. 
In this case, we can expand $k_4$ in powers of $\eta$,
\be
k_4(U-\epsilon P)= k_4(U)+k_4'(U) \epsilon P +k_4''(U) \frac{\epsilon^2 P^2}{2}+\cdots\,,
\ee
and write $D$ as
\be
D(P) = D_0+D_1  \epsilon P + D_2 \epsilon^2 P^2+\cdots\,,
\ee
where
\bea
D_0 &=&\alpha \frac{\pi i}{2} \int \frac{du_1}{u_1}\, h(u_1) \int dU k_1^*(U,0;u_1) \partial_U k_4\left(U,0;u_1\right)\,,\\
D_1 &=&  \alpha \frac{\pi i}{2} \int \frac{du_1}{u_1}\, h(u_1) \int dU k_1^*(U,0;u_1) \partial_U ^2k_4\left(U,0;u_1\right)\,,\\
D_2 &=&\frac{\alpha}{2} \frac{\pi i}{2} \int \frac{du_1}{u_1}\, h(u_1) \int dU k_1^*(U,0;u_1) \partial_U^3 k_4\left(U,0;u_1\right)\,.
\eea
Note that, for very small $\epsilon$, we are very close to the horizon, and we can write $D\approx D_0+D_1 \epsilon P$, which is very suggestive as it essentially reduces to the probe approximation. As we move the cutoff surface away from the horizon, we need to include higher order terms in (\ref{eq-Dseries}), consistent with the standard notion of UV/IR connection. At second order in $\epsilon P$, for example, we obtain the first backreaction effects, which were considered in Sec.~\ref{sec-beyondprobe}. Note that the infinite sum $\sum_{n=0}^{\infty} D_n (\epsilon P)^n$, corresponds to taking the cutoff to the AdS boundary, thus in this case we expect the exact function $D(P)$, which was computed, for example, in (\ref{eq-D(p)}). In this limit, however, the commutator naively diverges, since we are opening the wormhole with an irrelevant double trace deformation and $C$ is therefore very sensitive to the UV physics,

In Fig.~\ref{fig-DPnumerical}, we plot $D(P)$, as well as its second-order approximation $D_0+D_1 P+D_2 P^2$, for different values of the cutoff $r_c$. As we could have expected, the plot shows that as we move the cutoff closer and closer to the horizon ($r=r_0$), the second-order approximation for $D(P)$ becomes more and more precise. On the other hand, for sufficiently large values of $P$ the second-order approximation becomes unreliable, even when $r_c$ is very close to the horizon. We recall that the result from the second-order approximation was already qualitatively similar to the result obtained from smearing the operator, provided $P$ was not too big. So, if we want to guarantee that a small-$P$ approximation is reliable, it is natural to consider a smeared operator but now in the presence of a radial cutoff $r_c$.

\begin{figure}[t!]
\centering
\includegraphics[width=4in]{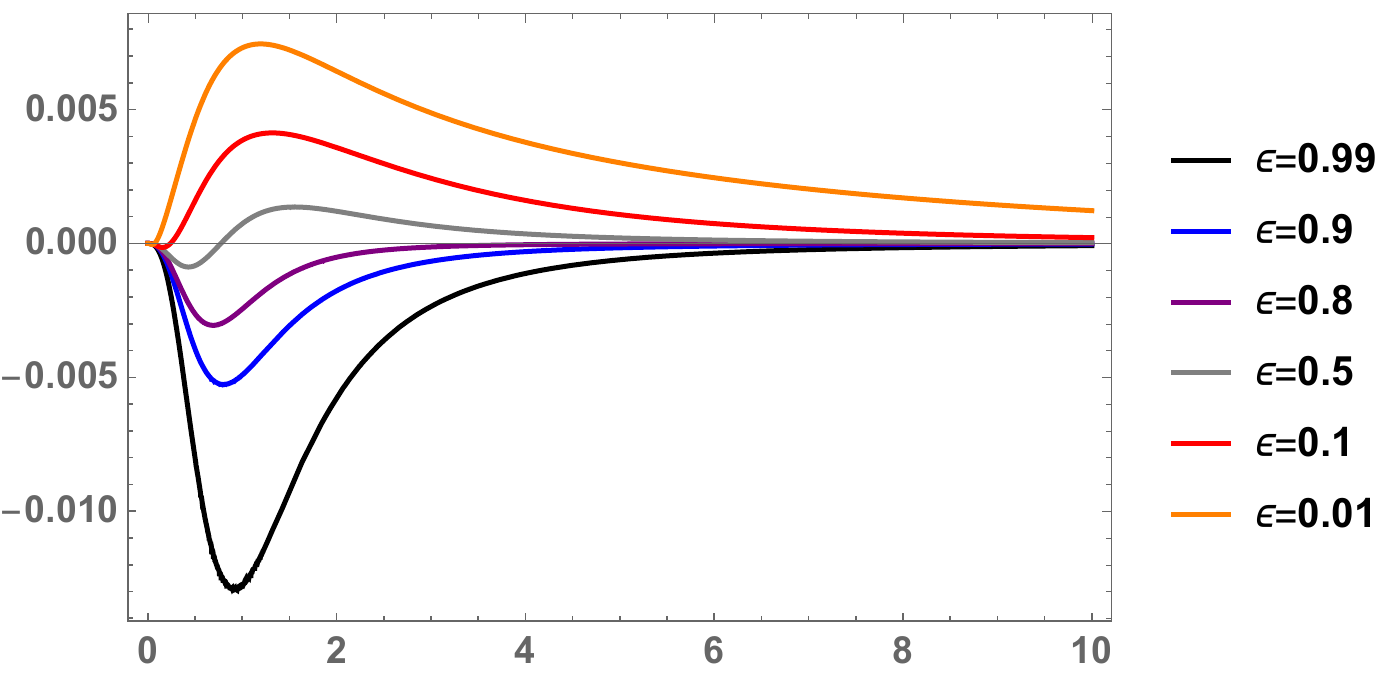}
\put(-50,0){$u_0$}
\put(-305,50){\rotatebox{90}{$D_1 = 2 \Delta v$}}

\caption{$D_1$ versus $u_0$ for different values of the cutoff $r_c$. Here $\epsilon = 1-r_0/r_c$, and $\Delta=1001/1000$. We set $\epsilon=0.99$ for the black curve, $\epsilon=0.9$ for the blue curve, $\epsilon=0.8$ for the purple curve, $\epsilon=0.5$ for the gray curve, $\epsilon=0.1$ for the red curve, and $\epsilon=0.01$ for the orange curve. Even though $D_1$ becomes positive for small values of $\epsilon$, the worldsheet wormhole can still be made traversable by changing the sign of $h$.  \label{fig-D1numerical}.}

\end{figure}

In Fig.~\ref{fig-D1numerical}, we plot $D_1$ vs $u_0$ for different values of the cutoff $r_c$.
Note that the sign of $D_1$ changes as we move the cutoff from a position close to the AdS boundary ($r_c=\infty$ or $\epsilon =1$) to  a position close to the horizon ($r_c=r_0$ or $ \epsilon =0$). When $D_1>0$, the wormhole can be made traversable by switching the sign of $h$, which also switches the sign of $D_1$.

\begin{figure}[t!]
    \centering
    \begin{minipage}{0.33\textwidth}
        \centering
        \includegraphics[width=0.9\textwidth]{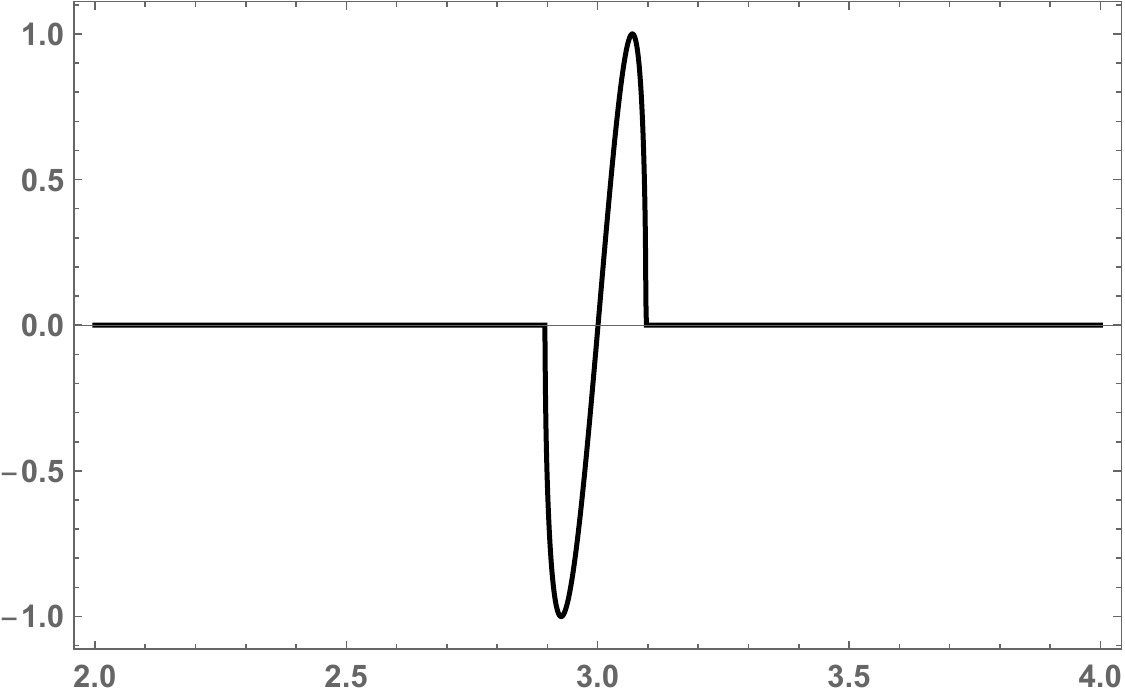} 
        \caption*{\small $r_c=10 \, r_0$}
    \end{minipage}\hfill
    \begin{minipage}{0.33\textwidth}
        \centering
        \includegraphics[width=0.9\textwidth]{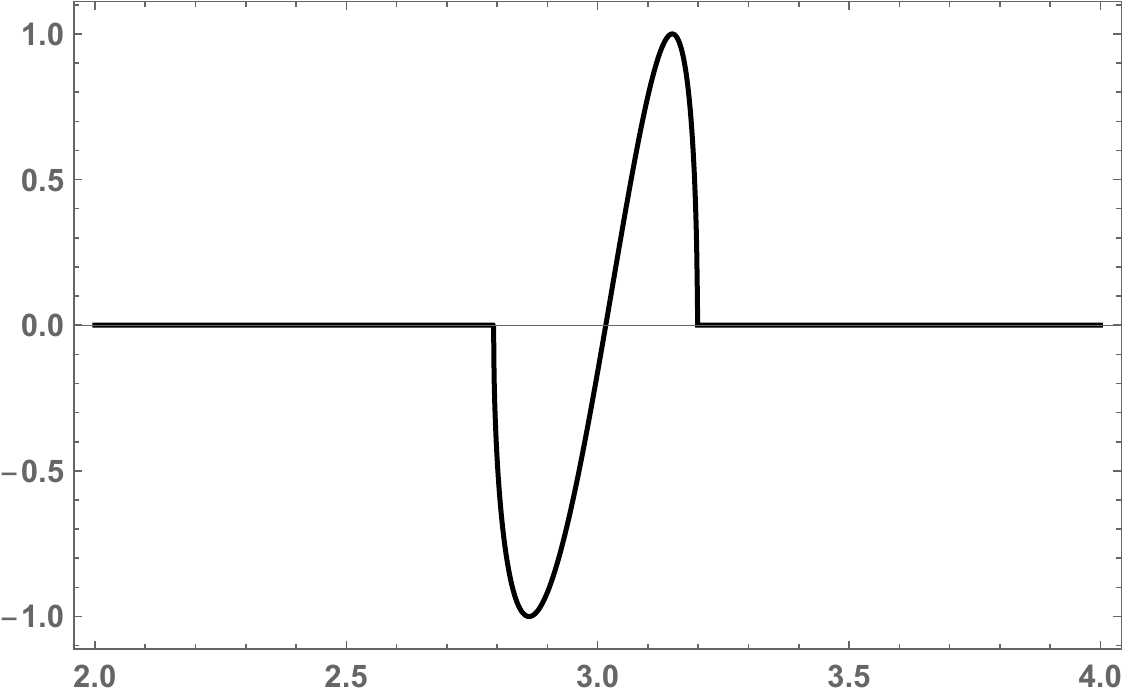} 
        \caption*{\small $r_c=5\, r_0$}
    \end{minipage}
    \begin{minipage}{0.33\textwidth}
        \centering
        \includegraphics[width=0.9\textwidth]{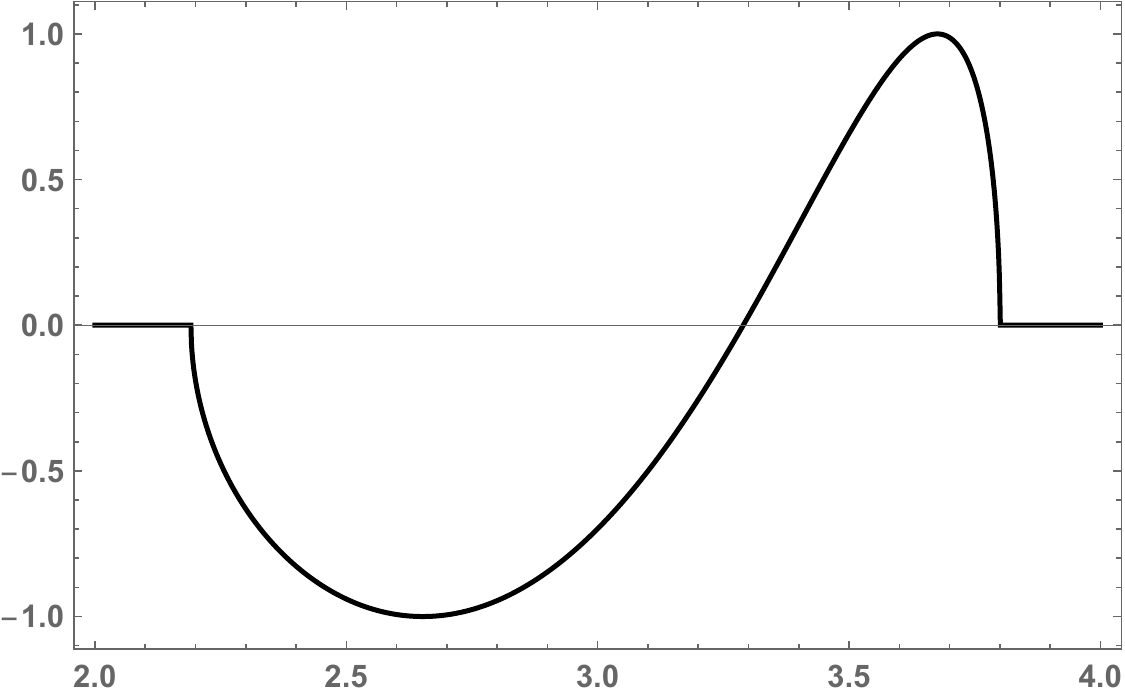} 
        \caption*{\small $r_c=1.5 \,r_0$}
    \end{minipage}
        \put(-445,-8){\rotatebox{90}{\small Im $C_\text{probe}$}}
        \put(-20,-35){\small time $t$}
    \caption{$\text{Im}\, C_\text{probe}(t;r_c)$ v.s. time $t$ for increasing values of $r_c/r_0$. Here we fixed $\alpha=-0.1$ and $\Delta_W=1$. \label{fig-Cs}}

\end{figure}

We now want to study and quantify the effects of the cutoff on the traversability. To do so, we employ a geodesic approximation to compute the two-sided correlator under the presence of a UV cutoff. The basic idea is that the double trace deformation  introduces a negative energy in the bulk, which opens the wormhole and shortens the distance between the two asymptotic boundaries. We compute the two-sided correlator as follows~\cite{Reynolds:2016pmi}
\be
\langle W_L(t_L) W_R(t_R) \rangle_{h} \sim e^{-\Delta_W d(t_L,t_R)}\,,
\ee
where $d(t_L,t_R)$ denotes the geodesic distance between the two boundary points, $t_L$ and $t_R$, and $\Delta_W$ is the scaling dimension of the operator $W$. In the above expression $\langle . \rangle_h$ denotes an expectation value  that is taken in the TFD state perturbed by the double trace deformation, i.e., in the presence of a negative energy shock. In embedding coordinates, the geodesic distance $d(P,P')$ between two bulk points $P=(T_1,T_2,X)$ and $P'=(T_1',T_2',X')$ is given by
\be
\cosh d(P,P') = T_1 T_1'+T_2 T_2'-X X'\,.
\ee
The embedding coordinates above are related to Kruskal $(u,v)$ and Rindler $(t,r)$ coordinates as follows
\bea
T_1 &=& \sqrt{r^2-1} \sinh t =\frac{u+v}{1+uv}\,,\\
T_2 &=& r =\frac{1-u v}{1+u v}\,,\\
X &=& \sqrt{r^2-1} \cosh t = \frac{v-u}{1+u v}\,,
\eea
where we have set $r_0=1$. Writing one of the points in Rindler coordinates, and the other in Kruskal coordinates, we find
\be
\cosh d(P,P')= \frac{1}{1+uv} \left[\sqrt{r^2-1}(u e^t-u e^{-t})+r(1-uv) \right]\,.
\ee
Following \cite{Shenker:2013pqa}, we write $d=\min_v ( d_1+d_2)$, where $d_1$ ($d_2$) denotes a distance between the left (right) asymptotic boundary and a point in the $u=0$ surface. One can show that
\bea
d_1&=& \text{arccosh}\left( {r_c-\sqrt{r_c^2-1}e^{-t_L} v} \right)\,,\\
d_2&=& \text{arccosh} \left( {r_c+\sqrt{r_c^2-1}e^{t_R} (v+\alpha)} \right)\,,
\eea
where $\alpha$ denotes the wormhole opening, and $r_c$ denotes the position of the cutoff. For $t_R=-t_L=t$, one finds that $v_\text{min}=-\alpha/2$. The two-sided correlator can then be written as
\bea \label{eq-Ccutoff}
\langle W_L(t_L) W_R(t_R) \rangle_{h} &\sim& e^{-\Delta_W (d_1+d_2)} \\
&=&\left[r_c+\frac{1}{2}\left( \alpha \, e^t \sqrt{r_c^2-1} +\sqrt{-2+2r_c+e^t\sqrt{r_c^2-1}}\sqrt{2+2r_c+e^t\sqrt{r_c^2-1}} \right) \right]^{-2\Delta_W}\!\!\!\!\!.\nonumber
\eea
In the limit $r_c \rightarrow \infty$, the above result reduces to $\left( 2+\alpha e^t \right)^{-2\Delta_W}$
which correspond to the probe limit result $C_\text{probe}$ obtained in previous sections with $t_R=-t_L=t$, and $\Delta v= \alpha$. This provides evidence that
\be
\text{Im} \left( \langle W_L(t_L) W_R(t_R) \rangle_{h} \right) \sim \langle [W_L(t_L),e^{-i \mathcal{V}}W_R(t_R)e^{i \mathcal{V}}] \rangle\,.
\ee
Going back to the case with finite $r_c$, we are interested in determining the time interval for which $\langle W_L(t_L) W_R(t_R) \rangle_{h}$ has a non-zero imaginary part, especially when we set $\Delta_W=1$. Studying (\ref{eq-Ccutoff}), one can show that $\langle W_L(t_L) W_R(t_R) \rangle_{h}$ becomes complex in the time interval
\be \label{eq-interval}
\frac{2}{|\alpha| }\sqrt{\frac{r_c-1}{r_c+1}} \leq \tilde{t} \leq \frac{2}{|\alpha| } \sqrt{\frac{r_c+1}{r_c-1}}\,,\,\,\,\,\,\, \tilde{t} \equiv e^t\,.
\ee
The above interval is centered around $t_c = \log \frac{2}{\alpha}$, which is the time scale that defines a sweet spot for traversability in the absence of a cutoff. The main effect of the cutoff is that now traversability is possible not only at $t=t_c$, but in a time window around $t_c$ that increases as move the cutoff surface deeper into the bulk (i.e., as $r_c$ approaches $r_0$). See Fig.~\ref{fig-Cs}. Moreover, the above result shows that the commutator is non-zero even when $\Delta_W$ takes integer values, which was not the case when $r_c \rightarrow \infty$ in the probe approximation.

\section{Discussion\label{sec:discussion}}

In this work, we studied worldsheet wormholes that become traversable after coupling the two endpoints of an open string in AdS. We considered the standard double trace deformation, $\delta\mathcal{L}\sim h \phi_L \phi_R$, introduced by Gao, Jafferis and Wall, as well as a second type of deformation of the form $\delta\mathcal{L}\sim h\partial\phi_L\partial\phi_R$, which seems more natural from the worldsheet perspective and could in principle emerge from the interactions between the string endpoints.

Normally, in higher dimensional setups, the traversability of wormholes can be diagnosed from violations of the ANEC, which states that the integral of the stress-energy tensor along complete achronal null geodesics is non-negative. Although this condition is satisfied in any local quantum field theory, the double-trace deformation that is introduced is non-local, invalidating one of the main assumptions of the ANEC and effectively rendering all null geodesics connecting the two boundaries non-achronal. Our two-dimensional setup includes a similar non-local coupling but does not contain gravity in the standard sense, so we needed to come up with a different criterion to check the traversability. In Sec.~\ref{sec-traversability}, we showed that worldsheet wormholes may indeed become traversable in the presence of worldsheet fluctuations with negative energy. More specifically, we showed that a signal originating in the left asymptotic boundary generally suffers a null shift $\Delta v$ in its trajectory, which is given by $\Delta v = \Delta E/8$, where $\Delta E$ is the energy of worldsheet fluctuations. This means that the signal may only reach the right asymptotic boundary if $\Delta v <0$, which requires $\Delta E <0$. This condition is analogous to the ANEC for standard gravitational setups and explains why traversable wormholes cannot arise when considering classical fluctuations of the string \cite{Chernicoff:2013iga}.

Using the point-splitting method, we computed the wormhole opening for two types of deformations and we found the following results:
\bea
\text{Case I:} \,\,\,\,\,\,\,\,\,\,\,\,\delta H_I &=& h \delta(t-t_0) \phi_L(-t_0)\phi_R(t_0) \, \nonumber \\
\Delta v^{\text{inst}}_{I}&=& -  \frac{h \ell_s^2}{2\pi}    \left( \frac{u_0}{1+u_0^2} \right)^{3}\\
\nonumber \\
\text{Case II:} \,\,\,\,\,\,\,\,\,\, \delta H_{II} &=& h \delta(t-t_0) \dot{\phi}_L(-t_0) \dot{\phi}_R(t_0) \, \nonumber \\
\Delta v^{\text{inst}}_{II}&=&  \frac{h \ell_s^2}{\pi}   \frac{u_0^{3}}{(1+u_0^2)^{5}} \Big[1-4 u_0^2+ u_0^4 \Big]
\eea
Using the eikonal approximation, we then computed the two-sided commutator
\be
\mathcal{C}(t_L,t_R) = \langle [\phi_L(t_L), e^{-i \mathcal{V}} \phi_R(t_R) e^{i \mathcal{V}}] \rangle
\ee
where $\mathcal{V}$ is a deformation of type I or II made with a large number of operators, $K$, defined as in (\ref{eq-deformationV}) \cite{Maldacena:2017axo}.\footnote{To achieve this we can consider a system of $K$ non-interacting strings, or a $K$-string. This approximation is useful because it enhances the effects of the coupling while suppressing particle creation in the worldsheet. However, $K$ cannot be too large, otherwise, interactions between individual strings become relevant. In such cases, a better description
of the system is in terms of a D3-brane carrying $K$ units of electric flux \cite{Drukker:2005kx,Gomis:2006im,Fiol:2014vqa}. It would be interesting to consider traversable wormholes in that setting.} This commutator measures how effectively we can send a message through the wormhole, displaying a non-zero value when the wormhole is traversable.
At the technical level, the calculation mimics the computation of the OTOC \cite{deBoer:2017xdk}, except that the operators $\mathcal{V}$ introduce negative energy as opposed to positive energy. Crucial to this calculation is the phase shift $\delta(s)$, which turns out to be linear in the Mandelstam variable $s=(\Delta E)^2$. This linear phase shift is surprising for a theory without dynamical gravitons\footnote{The linear shift arises from resummation of all possible higher spin excitations of the string.} and was singled out in \cite{deBoer:2017xdk} as responsible for maximal chaos. Our study hints at a more dramatic conclusion: \emph{wormhole traversability and maximal chaos go hand in hand, provided there is enough entanglement that can be used as a resource}.

Let us now discuss our results. First, we considered the signal in the probe approximation. In this case, the commutator was written in terms of the geodesic distance between the points $t_L$ and $t_R$ on the left and right asymptotic boundaries, respectively. The effect of the deformation of type I (type II) is to introduce null energy in the bulk, which in turn induces a decrease in the geodesic distance by $\Delta v_I$ ($\Delta v_{II}$). This commutator calculation thus provides an alternative way of computing the wormhole opening $\Delta v$, which  perfectly matched the result obtained by point-splitting. The first backreaction effects of the signal were also obtained analytically. The final result for the commutator with the first-order corrections was displayed in Fig.~\ref{fig-Cnonprobe}. We then studied numerically the full backreaction effects of the signal. In this case, however, it was important to consider a smeared version of the signal that suppresses high momentum contributions, since the operator deforming the theory is irrelevant. We analyzed two ways of smearing the operator, one by considering a hard cutoff and another one by adding an exponential damping term. Regardless of the way we cut the high-$P$ contributions, the results obtained were qualitatively similar to the results obtained by considering the first backreaction effects. This implies that the net effect of the irrelevant coupling could be accounted for by cutting off part of the asymptotic boundary, which effectively removes the aforementioned high-$P$ modes. This is in agreement with intuition from 2d dilaton gravity theories, where a similar effect also shows up when considering irrelevant deformations.

We can better understand the effects of the double-trace deformation on the near-boundary geometry if we think of the worldsheet as a $T \bar{T}$-like deformed theory  \cite{Cavaglia:2016oda}. The irrelevant deformation introduces a geometric cutoff $r_c$ that removes the asymptotic region of AdS, placing the dual boundary theory on a finite radial distance $r=r_c$ in the bulk \cite{McGough:2016lol}. With this interpretation in mind, we studied how the UV cutoff affects our results for the wormhole opening and for the commutator that diagnoses traversability. We observed that the UV cutoff improves the probe approximation and improves the traversability conditions, making it easier to send information through the wormhole. The appearance and value of the cutoff scale $r_c$ may be related to the strength of the irrelevant deformation $h$. To see this, notice that in the presence of the deformation, there is a scalar mode that grows near the boundary as $\phi\sim h\,r^{\Delta-d}$, and blows up for $\Delta>d$. To have a controlled theory, we need a UV cutoff such that $h\,r^{\Delta-d}<1$. As $h$ is increased, we are thus forced to pick smaller values of $r_c$ and eventually, let $r_c\to r_0$. Further, via the standard UV/IR relations in holography, one could map $r_c$ into a momentum scale $r_c\sim P_{\text{max}}$ in the dual theory, as part of the UV geometry is capped off. This is in complete agreement with our numerical findings concerning smeared operators.

An interesting future direction regarding worldsheet wormholes is the study of out-of-time-order six-point functions which characterize collisions in the wormhole interior. By introducing a coupling between the two asymptotic boundaries, one can make the worldsheet wormhole traversable and, thus, extract information about the collision product from behind the horizon \cite{Haehl:2021tft}. Another possible generalization would be to construct traversable wormholes on branes, instead of strings. This could be achieved by considering higher dimensional versions of the solutions used here \cite{Arcos:2022icf}. In fact, an attribute of worldvolume theories is that they may include dynamical gravity, in a way reminiscent of Randall-Sundrum braneworld models \cite{Randall:1999vf,Randall:1999ee}. In these setups, semiclassical backreaction may be treated exactly, i.e., including all orders in the Newton's constant $G_N$ (cf. \cite{Emparan:1999wa,Emparan:2002px,Emparan:2020znc,Emparan:2022ijy}). It would be interesting to investigate how the bounds on information transfer may be affected by semiclassical backreaction in these setups.

\noindent \noindent\section*{Acknowledgments}
We are grateful to Jose Barb\'on, Elena C\'aceres, Ben Freivogel, Alberto G\"uijosa, Arnab Kundu, Kyoungsun Lee, Ayan Patra, and Andrew Svesko for useful discussions and comments on the manuscript. JdB is supported by the European Research Council under the European Unions Seventh Framework Programme (FP7/2007-2013), ERC Grant agreement ADG 834878. VJ and KYK are supported by the National Research Foundation of Korea (NRF) funded by the Ministry of Education and the Ministry of Science, ICT \& Future Planning (NRF-2021R1A2C1006791 and NRF-
2020R1I1A1A01073135), the GIST Research Institute (GRI) and the AI-based GIST Research Scientist Project grant funded by the GIST in 2023. KYK is also supported by Creation of the Quantum Information Science R\&D Ecosystem (Grant No. 2022M3H3A106307411) through the National Research Foundation of Korea (NRF) funded by the Korean government (Ministry of Science and ICT). JFP is supported by the `Atracci\'on de Talento' program (2020-T1/TIC-20495, Comunidad de Madrid) and by the Spanish Research Agency (Agencia Estatal de Investigaci\'on) through the grants CEX2020-001007-S and PID2021-123017NB-I00, funded by MCIN/AEI/10.13039/501100011033 and by ERDF A way of making Europe.


\begin{thebibliography}{99}

\bibitem{Einstein:1935tc}
A.~Einstein and N.~Rosen,
``The Particle Problem in the General Theory of Relativity,''
Phys. Rev. \textbf{48}, 73-77 (1935).

\bibitem{Misner:1957mt}
C.~W.~Misner and J.~A.~Wheeler,
``Classical physics as geometry: Gravitation, electromagnetism, unquantized charge, and mass as properties of curved empty space,''
Annals Phys. \textbf{2}, 525-603 (1957).

\bibitem{Brown:2019hmk}
A.~R.~Brown, H.~Gharibyan, S.~Leichenauer, H.~W.~Lin, S.~Nezami, G.~Salton, L.~Susskind, B.~Swingle and M.~Walter,
``Quantum Gravity in the Lab: Teleportation by Size and Traversable Wormholes,''
[arXiv:1911.06314 [quant-ph]].

\bibitem{Nezami:2021yaq}
S.~Nezami, H.~W.~Lin, A.~R.~Brown, H.~Gharibyan, S.~Leichenauer, G.~Salton, L.~Susskind, B.~Swingle and M.~Walter,
``Quantum Gravity in the Lab: Teleportation by Size and Traversable Wormholes, Part II,''
[arXiv:2102.01064 [quant-ph]].

\bibitem{Maldacena:1997re}
J.~M.~Maldacena,
``The Large N limit of superconformal field theories and supergravity,''
Adv. Theor. Math. Phys. \textbf{2}, 231-252 (1998)
[arXiv:hep-th/9711200 [hep-th]].

\bibitem{Maldacena:2001kr}
J.~M.~Maldacena,
``Eternal black holes in anti-de Sitter,''
JHEP \textbf{04}, 021 (2003)
[arXiv:hep-th/0106112 [hep-th]].

\bibitem{VanRaamsdonk:2010pw}
M.~Van Raamsdonk,
``Building up spacetime with quantum entanglement,''
Gen. Rel. Grav. \textbf{42}, 2323-2329 (2010)
[arXiv:1005.3035 [hep-th]].

\bibitem{Maldacena:2013xja}
J.~Maldacena and L.~Susskind,
``Cool horizons for entangled black holes,''
Fortsch. Phys. \textbf{61}, 781-811 (2013)
[arXiv:1306.0533 [hep-th]].

\bibitem{Jensen:2013ora}
K.~Jensen and A.~Karch,
``Holographic Dual of an Einstein-Podolsky-Rosen Pair has a Wormhole,''
Phys. Rev. Lett. \textbf{111}, no.21, 211602 (2013)
[arXiv:1307.1132 [hep-th]].

\bibitem{Sonner:2013mba}
J.~Sonner,
``Holographic Schwinger Effect and the Geometry of Entanglement,''
Phys. Rev. Lett. \textbf{111}, no.21, 211603 (2013)
[arXiv:1307.6850 [hep-th]].

\bibitem{Chernicoff:2013iga}
M.~Chernicoff, A.~G\"uijosa and J.~F.~Pedraza,
``Holographic EPR Pairs, Wormholes and Radiation,''
JHEP \textbf{10}, 211 (2013)
[arXiv:1308.3695 [hep-th]].

\bibitem{Jensen:2014bpa}
K.~Jensen, A.~Karch and B.~Robinson,
``Holographic dual of a Hawking pair has a wormhole,''
Phys. Rev. D \textbf{90}, no.6, 064019 (2014)
[arXiv:1405.2065 [hep-th]].

\bibitem{Fischler:2014ama}
W.~Fischler, P.~H.~Nguyen, J.~F.~Pedraza and W.~Tangarife,
``Holographic Schwinger effect in de Sitter space,''
Phys. Rev. D \textbf{91}, no.8, 086015 (2015)
[arXiv:1411.1787 [hep-th]].

\bibitem{Hubeny:2014kma}
V.~E.~Hubeny and G.~W.~Semenoff,
``String worldsheet for accelerating quark,''
JHEP \textbf{10}, 071 (2015)
[arXiv:1410.1171 [hep-th]].

\bibitem{Xiao:2008nr}
B.~W.~Xiao,
``On the exact solution of the accelerating string in AdS(5) space,''
Phys. Lett. B \textbf{665}, 173-177 (2008)
[arXiv:0804.1343 [hep-th]].

\bibitem{Caceres:2010rm}
E.~Caceres, M.~Chernicoff, A.~Guijosa and J.~F.~Pedraza,
``Quantum Fluctuations and the Unruh Effect in Strongly-Coupled Conformal Field Theories,''
JHEP \textbf{06}, 078 (2010)
[arXiv:1003.5332 [hep-th]].


\bibitem{Atmaja:2010uu}
  A.~N.~Atmaja, J.~de Boer and M.~Shigemori,
  ``Holographic Brownian Motion and Time Scales in Strongly Coupled Plasmas,''
  Nucl.\ Phys.\ B {\bf 880}, 23 (2014)
  [arXiv:1002.2429 [hep-th]].

\bibitem{Banerjee:2013rca}
  P.~Banerjee and B.~Sathiapalan,
  ``Holographic Brownian Motion in 1+1 Dimensions,''
  Nucl.\ Phys.\ B {\bf 884}, 74 (2014)
  [arXiv:1308.3352 [hep-th]].


\bibitem{Shenker:2014cwa}
S.~H.~Shenker and D.~Stanford,
``Stringy effects in scrambling,''
JHEP \textbf{05}, 132 (2015)
[arXiv:1412.6087 [hep-th]].

\bibitem{Maldacena:2015waa}
J.~Maldacena, S.~H.~Shenker and D.~Stanford,
``A bound on chaos,''
JHEP \textbf{08}, 106 (2016)
[arXiv:1503.01409 [hep-th]].

\bibitem{Gao:2016bin}
P.~Gao, D.~L.~Jafferis and A.~C.~Wall,
``Traversable Wormholes via a Double Trace Deformation,''
JHEP \textbf{12}, 151 (2017)
[arXiv:1608.05687 [hep-th]].

\bibitem{Chernicoff:2009ff}
M.~Chernicoff, J.~A.~Garcia and A.~Guijosa,
``A Tail of a Quark in N=4 SYM,''
JHEP \textbf{09}, 080 (2009)
[arXiv:0906.1592 [hep-th]].


\bibitem{Maldacena:2017axo}
J.~Maldacena, D.~Stanford and Z.~Yang,
``Diving into traversable wormholes,''
Fortsch. Phys. \textbf{65}, no.5, 1700034 (2017)
[arXiv:1704.05333 [hep-th]].

\bibitem{Bak:2018txn}
D.~Bak, C.~Kim and S.~H.~Yi,
``Bulk view of teleportation and traversable wormholes,''
JHEP \textbf{08}, 140 (2018)
[arXiv:1805.12349 [hep-th]].

\bibitem{Bak:2019mjd}
D.~Bak, C.~Kim and S.~H.~Yi,
``Transparentizing Black Holes to Eternal Traversable Wormholes,''
JHEP \textbf{03}, 155 (2019)
[arXiv:1901.07679 [hep-th]].

\bibitem{Maldacena:2018gjk}
J.~Maldacena, A.~Milekhin and F.~Popov,
``Traversable wormholes in four dimensions,''
[arXiv:1807.04726 [hep-th]].

\bibitem{Ahn:2020csv}
B.~Ahn, Y.~Ahn, S.~E.~Bak, V.~Jahnke and K.~Y.~Kim,
``Holographic teleportation in higher dimensions,''
JHEP \textbf{07}, 219 (2021)
[arXiv:2011.13807 [hep-th]].

\bibitem{Ahn:2022rle}
B.~Ahn, S.~E.~Bak, V.~Jahnke and K.~Y.~Kim,
``Holographic teleportation with conservation laws: diffusion on traversable wormholes,''
[arXiv:2206.03434 [hep-th]].

\bibitem{Fu:2019vco}
Z.~Fu, B.~Grado-White and D.~Marolf,
``Traversable Asymptotically Flat Wormholes with Short Transit Times,''
Class. Quant. Grav. \textbf{36}, no.24, 245018 (2019)
[arXiv:1908.03273 [hep-th]].

\bibitem{AlBalushi:2020kso}
A.~Al Balushi, Z.~Wang and D.~Marolf,
``Traversability of Multi-Boundary Wormholes,''
JHEP \textbf{04}, 083 (2021)
[arXiv:2012.04635 [hep-th]].

\bibitem{Emparan:2020ldj}
R.~Emparan, B.~Grado-White, D.~Marolf and M.~Tomasevic,
``Multi-mouth Traversable Wormholes,''
JHEP \textbf{05}, 032 (2021)
[arXiv:2012.07821 [hep-th]].


\bibitem{Maldacena:2018lmt}
J.~Maldacena and X.~L.~Qi,
``Eternal traversable wormhole,''
[arXiv:1804.00491 [hep-th]].

\bibitem{Freivogel:2019lej}
B.~Freivogel, V.~Godet, E.~Morvan, J.~F.~Pedraza and A.~Rotundo,
``Lessons on eternal traversable wormholes in AdS,''
JHEP \textbf{07}, 122 (2019)
[arXiv:1903.05732 [hep-th]].

\bibitem{Bintanja:2021xfs}
S.~Bintanja, R.~Esp\'\i{}ndola, B.~Freivogel and D.~Nikolakopoulou,
``How to make traversable wormholes: eternal AdS$_{4}$ wormholes from coupled CFT\textquoteright{}s,''
JHEP \textbf{10}, 173 (2021)
[arXiv:2102.06628 [hep-th]].



\bibitem{Fallows:2020ugr}
S.~Fallows and S.~F.~Ross,
``Making near-extremal wormholes traversable,''
JHEP \textbf{12}, 044 (2020)
[arXiv:2008.07946 [hep-th]].

\bibitem{Caceres:2018ehr}
E.~Caceres, A.~S.~Misobuchi and M.~L.~Xiao,
``Rotating traversable wormholes in AdS,''
JHEP \textbf{12}, 005 (2018)
[arXiv:1807.07239 [hep-th]].




\bibitem{Maldacena:2020sxe}
J.~Maldacena and A.~Milekhin,
``Humanly traversable wormholes,''
Phys. Rev. D \textbf{103}, no.6, 066007 (2021)
[arXiv:2008.06618 [hep-th]].

\bibitem{Almheiri:2018ijj}
A.~Almheiri, A.~Mousatov and M.~Shyani,
``Escaping the Interiors of Pure Boundary-State Black Holes,''
[arXiv:1803.04434 [hep-th]].

\bibitem{Bao:2018msr}
N.~Bao, A.~Chatwin-Davies, J.~Pollack and G.~N.~Remmen,
``Traversable Wormholes as Quantum Channels: Exploring CFT Entanglement Structure and Channel Capacity in Holography,''
JHEP \textbf{11}, 071 (2018)
[arXiv:1808.05963 [hep-th]].

\bibitem{Freivogel:2019whb}
B.~Freivogel, D.~A.~Galante, D.~Nikolakopoulou and A.~Rotundo,
``Traversable wormholes in AdS and bounds on information transfer,''
JHEP \textbf{01}, 050 (2020)
[arXiv:1907.13140 [hep-th]].



\bibitem{Couch:2019zni}
J.~Couch, S.~Eccles, P.~Nguyen, B.~Swingle and S.~Xu,
``Speed of quantum information spreading in chaotic systems,''
Phys. Rev. B \textbf{102}, no.4, 045114 (2020)
[arXiv:1908.06993 [cond-mat.stat-mech]].

\bibitem{Marolf:2019ojx}
D.~Marolf and S.~McBride,
``Simple Perturbatively Traversable Wormholes from Bulk Fermions,''
JHEP \textbf{11}, 037 (2019)
[arXiv:1908.03998 [hep-th]].

\bibitem{Fu:2018oaq}
Z.~Fu, B.~Grado-White and D.~Marolf,
``A perturbative perspective on self-supporting wormholes,''
Class. Quant. Grav. \textbf{36}, no.4, 045006 (2019)
[erratum: Class. Quant. Grav. \textbf{36}, no.24, 249501 (2019)]
[arXiv:1807.07917 [hep-th]].



\bibitem{Hirano:2019ugo}
S.~Hirano, Y.~Lei and S.~van Leuven,
``Information Transfer and Black Hole Evaporation via Traversable BTZ Wormholes,''
JHEP \textbf{09}, 070 (2019)
[arXiv:1906.10715 [hep-th]].


\bibitem{Freivogel:2021ivu}
B.~Freivogel, D.~Nikolakopoulou and A.~F.~Rotundo,
``Wormholes from Averaging over States,''
[arXiv:2105.12771 [hep-th]].

\bibitem{Geng:2020kxh}
H.~Geng,
``Non-local entanglement and fast scrambling in de-Sitter holography,''
Annals Phys. \textbf{426}, 168402 (2021)
[arXiv:2005.00021 [hep-th]].

\bibitem{Nosaka:2020nuk}
T.~Nosaka and T.~Numasawa,
``Chaos exponents of SYK traversable wormholes,''
JHEP \textbf{02}, 150 (2021)
[arXiv:2009.10759 [hep-th]].

\bibitem{Levine:2020upy}
A.~Levine, A.~Shahbazi-Moghaddam and R.~M.~Soni,
``Seeing the entanglement wedge,''
JHEP \textbf{06}, 134 (2021)
[arXiv:2009.11305 [hep-th]].

\bibitem{Garcia-Garcia:2019poj}
A.~M.~Garc\'\i{}a-Garc\'\i{}a, T.~Nosaka, D.~Rosa and J.~J.~M.~Verbaarschot,
``Quantum chaos transition in a two-site Sachdev-Ye-Kitaev model dual to an eternal traversable wormhole,''
Phys. Rev. D \textbf{100}, no.2, 026002 (2019)
[arXiv:1901.06031 [hep-th]].

\bibitem{Numasawa:2020sty}
T.~Numasawa,
``Four coupled SYK models and nearly AdS$_{2}$ gravities: phase transitions in traversable wormholes and in bra-ket wormholes,''
Class. Quant. Grav. \textbf{39}, no.8, 084001 (2022)
[arXiv:2011.12962 [hep-th]].

\bibitem{Anand:2020wlk}
A.~Anand and P.~K.~Tripathy,
``Self-supporting wormholes with massive vector field,''
Phys. Rev. D \textbf{102}, 126016 (2020)
[arXiv:2008.10920 [hep-th]].

\bibitem{Anand:2022sbd}
A.~Anand,
``Self-Supporting Wormholes in Four Dimensions with Scalar Field,''
[arXiv:2204.08178 [hep-th]].


\bibitem{Kundu:2021nwp}
A.~Kundu,
``Wormholes and holography: an introduction,''
Eur. Phys. J. C \textbf{82}, no.5, 447 (2022)
[arXiv:2110.14958 [hep-th]].

\bibitem{Dubovsky:2012wk}
S.~Dubovsky, R.~Flauger and V.~Gorbenko,
``Solving the Simplest Theory of Quantum Gravity,''
JHEP \textbf{09}, 133 (2012)
[arXiv:1205.6805 [hep-th]].

\bibitem{Murata:2017rbp}
K.~Murata,
``Fast scrambling in holographic Einstein-Podolsky-Rosen pair,''
JHEP \textbf{11}, 049 (2017)
[arXiv:1708.09493 [hep-th]].

\bibitem{deBoer:2017xdk}
J.~de Boer, E.~Llabr\'es, J.~F.~Pedraza and D.~Vegh,
``Chaotic strings in AdS/CFT,''
Phys. Rev. Lett. \textbf{120}, no.20, 201604 (2018)
[arXiv:1709.01052 [hep-th]].

\bibitem{Banerjee:2018twd}
A.~Banerjee, A.~Kundu and R.~R.~Poojary,
``Strings, Branes, Schwarzian Action and Maximal Chaos,''
[arXiv:1809.02090 [hep-th]].

\bibitem{Banerjee:2018kwy}
A.~Banerjee, A.~Kundu and R.~Poojary,
``Maximal Chaos from Strings, Branes and Schwarzian Action,''
JHEP \textbf{06}, 076 (2019)
[arXiv:1811.04977 [hep-th]].

\bibitem{Vegh:2019any}
D.~Vegh,
``Celestial fields on the string and the Schwarzian action,''
JHEP \textbf{07}, 050 (2021)
[arXiv:1910.03610 [hep-th]].

\bibitem{Cavaglia:2016oda}
A.~Cavagli\`a, S.~Negro, I.~M.~Sz\'ecs\'enyi and R.~Tateo,
``$T \bar{T}$-deformed 2D Quantum Field Theories,''
JHEP \textbf{10}, 112 (2016)
[arXiv:1608.05534 [hep-th]].

\bibitem{Callebaut:2019omt}
N.~Callebaut, J.~Kruthoff and H.~Verlinde,
``$ T\overline{T} $ deformed CFT as a non-critical string,''
JHEP \textbf{04}, 084 (2020)
[arXiv:1910.13578 [hep-th]].

\bibitem{Chakraborty:2019mdf}
S.~Chakraborty, A.~Giveon and D.~Kutasov,
``$T\bar{T}$, $J\bar{T}$, $T\bar{J}$ and String Theory,''
J. Phys. A \textbf{52}, no.38, 384003 (2019)
[arXiv:1905.00051 [hep-th]].

\bibitem{McGough:2016lol}
L.~McGough, M.~Mezei and H.~Verlinde,
``Moving the CFT into the bulk with $ T\overline{T} $,''
JHEP \textbf{04}, 010 (2018)
[arXiv:1611.03470 [hep-th]].

\bibitem{Dubovsky:2017cnj}
S.~Dubovsky, V.~Gorbenko and M.~Mirbabayi,
``Asymptotic fragility, near AdS$_{2}$ holography and $ T\overline{T} $,''
JHEP \textbf{09}, 136 (2017)
[arXiv:1706.06604 [hep-th]].

\bibitem{Kraus:2018xrn}
P.~Kraus, J.~Liu and D.~Marolf,
``Cutoff AdS$_{3}$ versus the $ T\overline{T} $ deformation,''
JHEP \textbf{07}, 027 (2018)
[arXiv:1801.02714 [hep-th]].

\bibitem{Ferko:2022dpg}
C.~Ferko and S.~Sethi,
``Sequential Flows by Irrelevant Operators,''
[arXiv:2206.04787 [hep-th]].

\bibitem{Karch:2002sh}
A.~Karch and E.~Katz,
``Adding flavor to AdS / CFT,''
JHEP \textbf{06}, 043 (2002)
[arXiv:hep-th/0205236 [hep-th]].

\bibitem{Haag:1967sg}
R.~Haag, N.~M.~Hugenholtz and M.~Winnink,
``On the Equilibrium states in quantum statistical mechanics,''
Commun. Math. Phys. \textbf{5}, 215-236 (1967)

\bibitem{Gao:2018yzk}
P.~Gao and H.~Liu,
``Regenesis and quantum traversable wormholes,''
JHEP \textbf{10}, 048 (2019)
[arXiv:1810.01444 [hep-th]].

\bibitem{Reynolds:2016pmi}
A.~P.~Reynolds and S.~F.~Ross,
``Butterflies with rotation and charge,''
Class. Quant. Grav. \textbf{33}, no.21, 215008 (2016)
[arXiv:1604.04099 [hep-th]].

\bibitem{Shenker:2013pqa}
S.~H.~Shenker and D.~Stanford,
``Black holes and the butterfly effect,''
JHEP \textbf{03}, 067 (2014)
[arXiv:1306.0622 [hep-th]].

\bibitem{Drukker:2005kx}
N.~Drukker and B.~Fiol,
``All-genus calculation of Wilson loops using D-branes,''
JHEP \textbf{02}, 010 (2005)
[arXiv:hep-th/0501109 [hep-th]].

\bibitem{Gomis:2006im}
J.~Gomis and F.~Passerini,
``Wilson Loops as D3-Branes,''
JHEP \textbf{01}, 097 (2007)
[arXiv:hep-th/0612022 [hep-th]].

\bibitem{Fiol:2014vqa}
B.~Fiol, A.~G\"uijosa and J.~F.~Pedraza,
``Branes from Light: Embeddings and Energetics for Symmetric $k$-Quarks in $\mathcal{N}=4$ SYM,''
JHEP \textbf{01}, 149 (2015)
[arXiv:1410.0692 [hep-th]].

\bibitem{Haehl:2021tft}
F.~M.~Haehl, A.~Streicher and Y.~Zhao,
``Six-point functions and collisions in the black hole interior,''
JHEP \textbf{08}, 134 (2021)
[arXiv:2105.12755 [hep-th]].

\bibitem{Arcos:2022icf}
M.~Arcos, W.~Fischler, J.~F.~Pedraza and A.~Svesko,
``Membrane nucleation rates from holography,''
[arXiv:2207.06447 [hep-th]].

\bibitem{Randall:1999vf}
L.~Randall and R.~Sundrum,
``An Alternative to compactification,''
Phys. Rev. Lett. \textbf{83}, 4690-4693 (1999)
[arXiv:hep-th/9906064 [hep-th]].

\bibitem{Randall:1999ee}
L.~Randall and R.~Sundrum,
``A Large mass hierarchy from a small extra dimension,''
Phys. Rev. Lett. \textbf{83}, 3370-3373 (1999)
[arXiv:hep-ph/9905221 [hep-ph]].

\bibitem{Emparan:1999wa}
R.~Emparan, G.~T.~Horowitz and R.~C.~Myers,
``Exact description of black holes on branes,''
JHEP \textbf{01}, 007 (2000)
[arXiv:hep-th/9911043 [hep-th]].

\bibitem{Emparan:2002px}
R.~Emparan, A.~Fabbri and N.~Kaloper,
``Quantum black holes as holograms in AdS brane worlds,''
JHEP \textbf{08}, 043 (2002)
[arXiv:hep-th/0206155 [hep-th]].

\bibitem{Emparan:2020znc}
R.~Emparan, A.~M.~Frassino and B.~Way,
``Quantum BTZ black hole,''
JHEP \textbf{11}, 137 (2020)
[arXiv:2007.15999 [hep-th]].

\bibitem{Emparan:2022ijy}
R.~Emparan, J.~F.~Pedraza, A.~Svesko, M.~Toma\v{s}evi\'c and M.~R.~Visser,
``Black holes in dS$_{3}$,''
JHEP \textbf{11}, 073 (2022)
[arXiv:2207.03302 [hep-th]].

\end{thebibliography}
\end{document}